\documentclass[11pt,british]{article}
\usepackage[utf8]{inputenc}
\usepackage{geometry}
\usepackage{color}
\usepackage{babel}
\usepackage{booktabs}
\usepackage{mathtools}
\usepackage{amsmath}
\usepackage{amsthm}
\usepackage{amssymb}
\usepackage{graphicx}
\usepackage{setspace}
\usepackage{microtype}
\usepackage[unicode=true,
 bookmarks=true,bookmarksnumbered=false,bookmarksopen=false,
 breaklinks=false,pdfborder={0 0 0},pdfborderstyle={},backref=false,colorlinks=true]
 {hyperref}
\hypersetup{pdftitle={N=(2,0) AdS3 Solutions of M-theory},
 pdfauthor={Anthony Ashmore},
 colorlinks=true,urlcolor=darkblue,anchorcolor=darkblue,citecolor=darkblue,filecolor=darkblue,linkcolor=darkblue,menucolor=darkblue,linktocpage=true}

\makeatletter

\providecommand{\tabularnewline}{\\}

\usepackage{tikz}
\usetikzlibrary{backgrounds}
\usepackage[framemethod=tikz]{mdframed}
\AtBeginEnvironment{mdframed}{%
\tikzset{every picture/.style={}}%
}
\mdfsetup{roundcorner=.5ex,font=\small}
{\begin{mdframed}[backgroundcolor=white]}%
{\end{mdframed}}
{\begin{mdframed}[backgroundcolor=white]}%
{\end{mdframed}}

\usepackage{slashed}	 	
\usepackage{amsfonts} 		
\usepackage{cite}	
\SetTracking{encoding={*}, shape=sc}{0} 	
\allowdisplaybreaks		
\date{\today} 		
\numberwithin{equation}{section}	

\makeatletter
\g@addto@macro\bfseries{\boldmath}
\makeatother

\usepackage{xcolor}
\definecolor{darkblue}{rgb}{0.1,0.0,0.5}

\makeatletter
\g@addto@macro\@floatboxreset\centering
\makeatother

\let\originalleft\left
\let\originalright\right
\renewcommand{\left}{\mathopen{}\mathclose\bgroup\originalleft}
\renewcommand{\right}{\aftergroup\egroup\originalright}

\makeatother

\begin{document}
\global\long\def\rep#1{\boldsymbol{#1}}%
\global\long\def\repb#1{\overline{\boldsymbol{#1}}}%
\global\long\def\dd{\text{d}}%
\global\long\def\ii{\text{i}}%
\global\long\def\ee{\text{e}}%
\global\long\def\Dorf{L}%
\global\long\def\tDorf{\hat{L}}%
\global\long\def\GL#1{\text{GL}(#1)}%
\global\long\def\Orth#1{\text{O}(#1)}%
\global\long\def\SO#1{\text{SO}(#1)}%
\global\long\def\Spin#1{\text{Spin}(#1)}%
\global\long\def\Symp#1{\text{Sp}(#1)}%
\global\long\def\Uni#1{\text{U}(#1)}%
\global\long\def\SU#1{\text{SU}(#1)}%
\global\long\def\Gx#1{\text{G}_{#1}}%
\global\long\def\Fx#1{\text{F}_{#1}}%
\global\long\def\Ex#1{\text{E}_{#1}}%
\global\long\def\ExR#1{\text{E}_{#1}\times\mathbb{R}^{+}}%
\global\long\def\ex#1{\mathfrak{e}_{#1}}%
\global\long\def\gl#1{\mathfrak{gl}_{#1}}%
\global\long\def\SL#1{\text{SL}(#1)}%
\global\long\def\Stab{\operatorname{Stab}}%
\global\long\def\vol{\operatorname{vol}}%
\global\long\def\tr{\operatorname{tr}}%
\global\long\def\ad{\operatorname{ad}}%
\global\long\def\ext{\Lambda}%
\global\long\def\AdS#1{\text{AdS}_{#1}}%
\global\long\def\op#1{\operatorname{#1}}%
\global\long\def\im{\operatorname{im}}%
\global\long\def\re{\operatorname{re}}%
\global\long\def\eqspace{\mathrel{\phantom{{=}}{}}}%
\global\long\def\bZ{\mathbb{Z}}%
\global\long\def\bC{\mathbb{C}}%
\global\long\def\bP{\mathbb{P}}%
\global\long\def\bR{\mathbb{R}}%
\global\long\def\feyn#1{\slashed{#1}}%
\global\long\def\id{\operatorname{id}}%
\global\long\def\ap{\alpha'}%
\global\long\def\del{\partial}%
\global\long\def\bdel{\bar{\partial}}%
\global\long\def\transpose{{\scriptscriptstyle \mathsf{T}}}%
\global\long\def\charge{\text{c}}%
\let\oldstar\star\renewcommand{\star}{\mathop{}\mathopen{}{\oldstar}}

\begin{titlepage}
\begin{flushright} 
\end{flushright}
\vfill
\begin{center}
{\setstretch{1.3}\Large\bf $N=(2,0)$ AdS$_3$ Solutions of M-theory\par} 
\vskip 1cm 
Anthony Ashmore
\vskip 1cm
\textit{\small{}Enrico Fermi Institute \& Kadanoff Center for Theoretical Physics,\\ University of Chicago, Chicago, IL 60637, USA}
\end{center}
\vfill
\begin{center} \textbf{Abstract} \end{center}
\begin{quote}
We consider the most general solutions of eleven-dimensional supergravity preserving $N=2$ supersymmetry whose metrics are warped products of three-dimensional anti-de Sitter space with an eight-dimensional manifold, focusing on those realising (2,0) superconformal symmetry. We give a set of necessary and sufficient conditions for a solution to be supersymmetric, which can be phrased, in the general case, in terms of a local SU(2) structure and its intrinsic torsion. We show that these supergravity backgrounds always admit a nowhere-vanishing Killing vector field that preserves the solution and encodes the U(1) R-symmetry of the dual field theory. We illustrate our results with examples which have appeared in the literature, including those with SU(4), G$_2$ and SU(3) structures, and discuss new classes of Minkowski solutions.
\end{quote}
\vfill
{\begin{NoHyper}\let\thefootnote\relax\footnotetext{\tt \!\!\!\!\!\!\!\!\!\!\!\! ashmore@uchicago.edu}\end{NoHyper}}
\end{titlepage}

\microtypesetup{protrusion=false} 
\tableofcontents 
\microtypesetup{protrusion=true} 

\section{Introduction}

Supersymmetric AdS$_{3}$ backgrounds in string and M-theory provide a rich and relatively unexplored arena where supergravity and holography can be studied. Of particular interest to the author are two observations. First, the two-dimensional conformal field theories dual to such backgrounds are often better understood than their higher-dimensional counterparts, particularly with powerful results for chiral $N=2$ theories such as $c$-extremisation~\cite{1211.4030,1302.4451}. For theories with type IIB duals, the gravitational version of this extremisation principle is understood for backgrounds with only five-form flux~\cite{1810.11026,1812.05597,1901.05977,1910.08078}. Since $c$-extremisation is independent of the details of the dual gravitational solution, there should be an as-yet-unknown extremisation principle applicable to general AdS$_{3}$ backgrounds in string or M-theory preserving $N=(2,0)$ supersymmetry. Secondly, the KKLT construction~\cite{Kachru:2003aw} of de Sitter backgrounds in string theory relies on a landscape of supersymmetric AdS$_{4}$ vacua with small cosmological constant~\cite{Giddings:2001yu}. It is somewhat unclear if current results can provide such a landscape or whether the solutions needed actually exist at all~\cite{Sethi:2017phn,Bena:2018fqc,Obied:2018sgi,Danielsson:2018ztv,Gautason:2018gln,Danielsson:2018qpa,Gao:2020xqh} As an alternative to tackling this problem head-on, one might wonder whether there is a similar landscape of supersymmetric $\AdS 3$ backgrounds.

For both of these questions, we need a thorough understanding of the geometry underlying the supergravity backgrounds, in particular the conditions for preserving supersymmetry. In this paper, we fill a gap in the literature by analysing the most general AdS$_{3}$ solutions to eleven-dimensional supergravity preserving $N=(2,0)$ supersymmetry.\footnote{The distinction between $(2,0)$ and $(0,2)$ is a matter of convention.  In particular, some of the following papers refer to $(0,2)$ and $(0,4)$ supersymmetry, which we refer to as $(2,0)$ and $(4,0)$ respectively.}

There are many previous works studying supersymmetric AdS$_{3}$ backgrounds in both string and M-theory. For type IIB, there are known conditions for backgrounds preserving $(1,0)$ supersymmetry \cite{1910.06326,Passias:2020ubv}, $(1,1)$ supersymmetry~\cite{Macpherson:2021lbr} $(2,0)$ backgrounds \cite{Kim:2005ez,0807.4375,1712.07631} with the general conditions given in \cite{1911.04439}, $(2,2)$ backgrounds with pure NS-NS flux \cite{1710.09826} or pure five-form flux \cite{Couzens:2021tnv}, and $(4,0)$ backgrounds \cite{Couzens:2017way,Macpherson:2022sbs}. There are now a great many solutions which fall within these classifications -- see \cite{hep-th/0612253,hep-th/0606221,1511.09462,1912.07605,1402.3807} and references therein. There is also a rich collection of results in type IIA \cite{1812.10172,1807.06602,Passias:2020ubv}, massive IIA \cite{1909.11669,1909.10510,1908.09851,1909.09636,Lozano:2015bra,Lozano:2022ouq,Couzens:2022agr}, and heterotic supergravity \cite{1505.01693}. Important for us are the results on Minkowski or AdS$_{3}$ backgrounds in eleven-dimensional supergravity preserving $N=1$ supersymmetry \cite{hep-th/0011114,hep-th/0201227,Martelli:2003ki,hep-th/0511047},\footnote{The geometry of the foliated eight-manifolds in the $N=1$ case is studied in detail in \cite{Babalic:2014fua,Babalic:2014dea,Babalic:2015kpa,Shahbazi:2015mva}.} $N=2$ supersymmetry \cite{hep-th/9605053,hep-th/9908088}, $N=4$ supersymmetry~\cite{1005.4527,Kim:2007hv,Lozano:2015bra,Kelekci:2016uqv,Faedo:2020nol,Lozano:2020bxo,Dibitetto:2020bsh}, and $N=8$ supersymmetry~\cite{DHoker:2008lup,Estes:2012vm,Bachas:2013vza,Legramandi:2020txf}. 

Previous results for $(2,0)$ backgrounds of M-theory on eight-manifolds have appeared in the literature, focusing on ``geometric algebra'' techniques~\cite{Lazaroiu:2012du}. In particular, the work of \cite{1212.6918,1301.5106,Babalic:2013iha,1411.3493} gives a set of differential and algebraic constraints on objects that live on a metric cone over the eight-manifold. In the physics literature, \cite{1311.5901} studied compactifications with local $\SU 3$ structures for which probe M2-branes filling the spacetime feel a potential.\footnote{See also \cite{Micu:2016bmz} for related work.} The results of \cite{Martelli:2003ki} include an $N=2$ Minkowski background with a global $\SU 3$ structure, while work on wrapped branes has led to a partial classification of backgrounds with $\SU 3$ structures~\cite{0907.1665,hep-th/0605146,hep-th/0608055,1508.04135}, with recent applications including M5-branes wrapping four-dimensional orbifolds~\cite{2204.02990}.\footnote{See also \cite{Figueras:2007cn} for a discussion of the global geometry of supersymmetric $\AdS 3$ solutions in eleven-dimensional supergravity.} However, a general $G$-structure analysis and a study of the geometry and physics of the resulting backgrounds has been missing.

In this work, we derive the necessary and sufficient conditions for AdS$_{3}$ backgrounds of eleven-dimensional supergravity to preserve $(2,0)$ supersymmetry without any assumptions on the fluxes. Using the $G$-structure formalism \cite{hep-th/0205050,Gauntlett:2003cy}, we give these conditions in terms of differential and algebraic conditions on a set of differential forms defined from spinor bilinears. We show there always exists a nowhere-vanishing Killing vector on the internal space, dual to the R-symmetry of the $(2,0)$ SCFT, and discuss when the Bianchi identities and equations of motion are implied by supersymmetry alone. Generically, the internal eight-manifold admits a local $\SU 2$ structure, which can enhance to $\SU 3$, $\Gx 2$ or $\SU 4$. We also consider special cases where the eight-manifold admits a global $\SU 4$, $\SU 3$ or $\Gx 2$ structure. In each case, we analyse the simplified supersymmetry conditions and make contact with known solutions in the literature. A pair of Mathematica notebooks with the Killing spinor and $G$-structure calculations is included with the submission.

The paper is organised as follows. In Section \ref{sec:Conditions-for-}, we discuss the conditions for an $N=(2,0)$ AdS$_{3}$ background in eleven-dimensional supergravity in terms of Killing spinors, describe how the spinor bilinears define a set of $p$-forms, and give the differential and algebraic conditions that correspond to the Killing spinor equations. We also show how and when supersymmetry implies the Bianchi identities and equations of motion, and discuss the existence of a nowhere-vanishing Killing vector, dual to the R-symmetry of the $(2,0)$ field theory. In Section \ref{sec:Analysis-of--structures}, we review results in the literature that constrain the local $G$-structure on the internal eight-manifold, focusing on how the values of certain scalars determine whether the pair of Killing spinors is stabilised by $\SU 2$, $\SU 3$, $\Gx 2$ or $\SU 4$. In Section \ref{sec:Examples}, we use our results to reproduce a number of known solutions, including Calabi--Yau fourfolds with electric flux, $\Gx 2$ holonomy manifolds, branes wrapping special Lagrangian three-cycles in $\SU 3$ structure manifolds, and AdS backgrounds from branes wrapping Kähler or co-associative four-cycles. We also give some preliminary results on backgrounds which admit generic local $\SU 3$ structures or a certain local $\SU 2$ structure. We finish with some discussion in Section \ref{sec:Conclusions-and-outlook}, and relegate our conventions and an orthonormal frame for the $G$-structures to the appendices.

\section{Conditions for \texorpdfstring{$N=2$}{N = 2} supersymmetry in eight dimensions\label{sec:Conditions-for-}}

We begin with a review of the most general ansatz for the metric and flux compatible with the isometries of AdS$_{3}$, and then present the Killing spinor equations from which our analysis follows. We then translate the existence of two independent solutions to these equations into differential and algebraic conditions on a set of spinor bilinears which characterise a local $G$-structure on the internal eight-manifold. We discuss which Bianchi identities and supergravity equations of motion are automatically implied by supersymmetry. As for $(2,0)$ backgrounds in type IIB supergravity \cite{1911.04439}, one expects the existence of a Killing vector field -- known as the \emph{R-symmetry vector} -- which preserves the full solution (metric, warp factor and flux) and is dual to the R-symmetry of the two-dimensional SCFT. We end by showing that such a Killing vector always exists for AdS$_{3}$ solutions preserving $(2,0)$ supersymmetry.

\subsection{Eleven-dimensional supergravity on eight-manifolds}

The field content of eleven-dimensional supergravity is a metric, a local three-form $C$ with field strength $G=\dd C$, and a gravitino $\Psi_{M}$. We follow the conventions of \cite{Martelli:2003ki}. A supersymmetric background of this theory is a solution with vanishing gravitino admitting (at least) one nowhere-vanishing Majorana spinor $\eta$ for which the gravitino variation vanishes:
\begin{equation}
\delta\Psi_{M}\equiv\hat{\nabla}_{M}\eta-\tfrac{1}{288}\left(\hat{\Gamma}^{NPQR}{}_{M}G_{NPQR}-8G_{MNPQ}\hat{\Gamma}^{NPQ}\right)\eta=0,\label{eq:gravitino_variation}
\end{equation}
where the $\hat{\Gamma}_{M}$ furnish a representation of the Clifford algebra in eleven dimensions. We further require the eleven-dimensional geometry to be Poincar\'e or AdS invariant in three external dimensions, and thus take the background to be a warped product of three-dimensional Minkowski or AdS space with an eight-manifold $X$. The metric then splits as
\begin{equation}
\dd s_{11}^{2}=\ee^{2\Delta}\left(\dd s^{2}(\AdS 3)+\dd s^{2}(X)\right),\label{eq:11_metric}
\end{equation}
where $\Delta$ is a warp factor that depends only on the internal coordinates. The most general four-form field strength $G$ compatible with this ansatz is
\begin{equation}
G=\ee^{3\Delta}(F+\vol_{3}\wedge f),\label{eq:flux_warp}
\end{equation}
where $\vol_{3}$ is the volume form of AdS$_{3}$, $F$ is a four-form on $X$, and $f$ is a one-form on $X$. We refer to $F$ and $f$ as the magnetic and electric components of the flux respectively. The equation of motion and Bianchi identity for $G$ are then equivalent to
\begin{align}
\dd(\ee^{3\Delta}F) & =0, & \ee^{-6\Delta}\dd(\ee^{6\Delta}\star f)+\tfrac{1}{2}F\wedge F & =0,\label{eq:Bianchi (1,0)}\\
\dd(\ee^{3\Delta}f) & =0, & \ee^{-6\Delta}\dd(\ee^{6\Delta}\star F)-f\wedge F & =0,\label{eq:implied}
\end{align}
where $\star$ is the Hodge star associated to the unwarped metric on the eight-manifold $X$.

\subsection{Spinor ansatz}

We want to analyse the supersymmetry conditions for a general $N=(2,0)$ background. Such backgrounds admit two linearly independent solutions to vanishing of the gravitino variation \eqref{eq:gravitino_variation}. Recall that an eleven-dimensional Majorana spinor can be decomposed as
\begin{equation}
\eta=\ee^{\Delta/2}\psi\otimes\chi,\label{eq:N=00003D1_ansatz}
\end{equation}
where $\eta$, $\psi$ and $\chi$ are spinors of $\Spin{1,10}$, $\Spin{1,2}$ and $\Spin 8$ respectively. Given our conventions in Appendix \ref{app:conventions} where the eight-dimensional gamma matrices are real and symmetric, Majorana spinors in eight dimensions are simply real spinors, so the Majorana condition on $\eta$ amounts to
\begin{equation}
\psi^{*}=\gamma_{2}\psi,\qquad\chi^{*}=\chi.
\end{equation}
The general $N=2$ spinor ansatz can then be written as
\begin{equation}
\eta=\ee^{\Delta/2}\sum_{i=1,2}\psi_{i}\otimes\chi_{i},
\end{equation}
where the $\chi_{i}$ are Majorana and nowhere-vanishing on $X$. For an $N=1$ background with a three-dimensional Minkowski or AdS factor, we require that the external spinor $\psi$ satisfies
\begin{equation}
\nabla_{\mu}\psi+m\gamma_{\mu}\psi=0.\label{eq:ads_killing}
\end{equation}
In the $N=2$ case, we take both $\psi_{i}$ to satisfy the AdS$_{3}$ Killing spinor equation
\begin{equation}
\nabla_{\mu}\psi_{i}+m_{i}\gamma_{\mu}\psi_{i}=0,
\end{equation}
where $m_{1}=m$ and $m_{2}=\pm m$. Here $m$ is proportional to the inverse AdS radius and normalised so that the Ricci scalar for AdS$_{3}$ is $R=-24m^{2}$. For Minkowski backgrounds with $m=0$, each internal spinor leads to two real supercharges, so that the $N=2$ ansatz gives four real supercharges, as expected. For AdS$_{3}$ backgrounds, there is a further refinement since the supersymmetry algebra factorises into left and right R-symmetries~\cite{Nahm:1977tg,Gunaydin:1986fe}. For $m_{2}=+m$, the supersymmetry is chiral, which we write as $N=(2,0)$.\footnote{Since the sign of $m$ is not fixed, what we call $(2,0)$ also parametrises $(0,2)$.} When $X$ is compact, the dual CFT living on the AdS boundary then has $(2,0)$ supersymmetry in two dimensions.\footnote{See \cite[Appendix B]{Couzens:2017way} for a lucid discussion of how AdS$_{3}$ supercharges relate to the preserved supersymmetries in the dual two-dimensional SCFT.} For $m_{2}=-m$, one has a non-chiral $(1,1)$ supersymmetry. In what follows, we focus on $(2,0)$ backgrounds and thus take $m_{1}=m_{2}=m$.

In the $(2,0)$ case, the boundary SCFT should admit a $\Uni 1$ R-symmetry. As we show in Section \ref{subsec:A-Killing-vector}, in the dual geometry, this $\Uni 1$ appears as a rotation of the spinors as a doublet. To see this, it is convenient to define the complex spinors
\begin{equation}
\chi_{\pm}=\frac{1}{\sqrt{2}}(\chi_{1}\pm\ii\chi_{2}),
\end{equation}
with the $\Uni 1$ acting as a phase. Note that, since the $\chi_{i}$ are real, we have $\chi_{\pm}^{\charge}=\chi_{\mp}$. To simplify our notation somewhat, we denote $\chi=\chi_{+}$ from here onwards, with $\chi^{\charge}=\chi_{-}$. We can then write the $N=2$ spinor ansatz as
\begin{equation}
\ee^{-\Delta/2}\eta=\psi_{-}\otimes\chi+\psi_{+}\otimes\chi^{\charge},
\end{equation}
where we have defined the complex external spinors
\begin{equation}
\psi_{\pm}=\frac{1}{\sqrt{2}}(\psi^{1}\pm\ii\psi^{2}).
\end{equation}
These spinors transform with a phase under the $\Uni 1$ R-symmetry and become the supercharges in the SCFT on the boundary of AdS$_{3}$.

\subsection{Killing spinor equations}

As in \cite{Martelli:2003ki}, reducing the gravitino variation \eqref{eq:gravitino_variation} with the ansatz \eqref{eq:N=00003D1_ansatz} gives internal and external equations in terms of the internal spinors. For $N=2$ supersymmetry, we need two independent solutions $\chi_{1}$ and $\chi_{2}$ to these equations. In terms of the complex spinor $\chi$, these equations are\footnote{Note that this agrees with the comment in \cite{Martelli:2003ki} that the $N=(2,0)$ case follows from taking the external and internal spinors to be complex. This is not true for the $N=(1,1)$ case as the Killing spinor equations mix $\chi_{+}$ and $\chi_{-}$.}
\begin{align}
0 & =\nabla_{m}\chi+\tfrac{1}{24}F_{mpqr}\gamma^{pqr}\chi-\tfrac{1}{4}f_{n}\gamma^{n}{}_{m}\gamma_{9}\chi+m\gamma_{m}\gamma_{9}\chi,\label{eq:KS_eq}\\
0 & =\tfrac{1}{2}\partial_{m}\Delta\gamma^{m}\chi-\tfrac{1}{288}F_{mnpq}\gamma^{mnpq}\chi-\tfrac{1}{6}f_{m}\gamma^{m}\gamma_{9}\chi-m\gamma_{9}\chi,\label{eq:KS_eq_alg}
\end{align}
where $\gamma_{9}=\gamma_{1}\dots\gamma_{8}$ is the highest-rank gamma matrix.  We refer these collectively as the Killing spinor equations, and often call the upper / lower equation the differential / algebraic Killing spinor equation. Solutions to these equations are often known as Killing spinors.

\subsection{Local \texorpdfstring{$G$}{G}-structures and spinor bilinears}

It is now well established that the formalism of $G$-structures and their intrinsic torsion gives a powerful tool for understanding supersymmetric flux backgrounds. In this paper, it is useful to distinguish between local and global $G$-structures. A local $G$-structure on $X$ is a set of locally defined $G$-invariant objects transforming in irreducible representations of $\Spin 8$, where $G\subset\Spin 8$. Since these objects may only be locally defined, and the differential conditions they satisfy need be defined only in an open set, we refer to this as a local $G$-structure.\footnote{As noted in \cite{1505.02270}, there is a discrepancy between the use of the term ``local $G$-structure'' in mathematics and physics. We use the physics terminology where a ``local'' structure holds for points in some open set, which itself might be a closed subset of the eight-manifold $X$ (for example, on a lower-dimensional subspace of $X$). These are known as ``stratified'' structures in the mathematics literature.} If the invariant objects can be extended globally over $X$, they define a $G$-structure in the usual sense, i.e.~a principal $G$-subbundle of the frame bundle of $X$. Where emphasis is necessary, we refer to this stronger notion as a global $G$-structure. As we discuss in detail in Section \ref{sec:Analysis-of--structures}, the nowhere-vanishing spinors $(\chi_{1},\chi_{2})$ are generically stabilised by $\SU 2$. Along certain loci on $X$ or for certain choices of scalar parameters, the stabiliser can enhance to $\SU 3$, $\Gx 2$ or $\SU 4$.

Since the Majorana spinors $\chi_{i}$ are real and the Clifford algebra intertwiners are trivial, the scalar bilinears are $\chi_{i}\chi_{j}$ and $\chi_{i}\gamma_{9}\chi_{j}$, or equivalently $\bar{\chi}\chi$ and $\bar{\chi}^{\charge}\chi$. The differential Killing spinor equation \eqref{eq:KS_eq} immediately implies
\begin{equation}
\nabla(\chi_{1}\chi_{1})=\nabla(\chi_{2}\chi_{2})=0,
\end{equation}
and so the norms of $\chi_{1}$ and $\chi_{2}$ are constant. Clearly, this means that for $N=(2,0)$ backgrounds $\chi_{1}$ and $\chi_{2}$ are nowhere-vanishing, as we expected. Since the Killing spinor equations are invariant under constant $\GL{2,\bR}$ transformations of $(\chi_{1},\chi_{2})$, we can use $\bR^{*}\times\bR^{*}\subset\GL{2,\bR}$ to set $\chi_{1}\chi_{1}=\chi_{2}\chi_{2}=1$ without loss of generality. One can also use \eqref{eq:KS_eq} to show $\nabla(\chi_{1}\chi_{2})=0$, and so $\chi_{1}\chi_{2}$ is also constant. Since the Majorana spinors $\chi_{i}$ are real, we can use the remaining $\GL{2,\bR}$ freedom to set $\chi_{1}\chi_{2}=0$. In the $\chi_{\pm}$ basis, these choices correspond to
\begin{equation}
\bar{\chi}\chi=\chi\chi^{\charge}=1,\qquad\bar{\chi}^{\charge}\chi=\bar{\chi}\chi^{\charge}=0.\label{eq:norm_orthogonal}
\end{equation}

As in the $N=1$ case of \cite{Martelli:2003ki}, one finds that the scalars containing $\gamma_{9}$ are not constant in general, $\nabla(\chi_{i}\gamma_{9}\chi_{j})\neq0$. There are three independent scalar bilinears that one can define, namely $\chi_{1}\gamma_{9}\chi_{1}$, $\chi_{2}\gamma_{9}\chi_{2}$ and $\chi_{1}\gamma_{9}\chi_{2}$. In the complex spinor basis, we can capture these as
\begin{equation}
\bar{\chi}\gamma_{9}\chi\equiv\zeta,\qquad\bar{\chi}^{\charge}\gamma_{9}\chi\equiv S,\label{eq:zeta_S_def}
\end{equation}
where $\zeta$ is real and $S$ is complex. Given the normalisation of $\chi_{1}$ and $\chi_{2}$, the Cauchy--Schwarz inequality implies $|\zeta|\leq1$ and $|S|\leq1$. Furthermore, as we review in Section \ref{sec:Analysis-of--structures}, these scalars obey
\begin{equation}
\zeta^{2}+|S|^{2}\leq1.\label{eq:scalar_norm}
\end{equation}
One can then parametrise the scalars as
\begin{equation}
\zeta=\sin\alpha\cos\delta,\qquad S=\ee^{\ii\sigma}\sin\alpha\sin\delta,\label{eq:S_zeta_def}
\end{equation}
where $\alpha\in[0,\pi/2]$ determines the value of $\zeta^{2}+|S|^{2}$. We use both of these parametrisations in what follows.

For each rank $n$, we have the following bilinears
\begin{equation}
\bar{\chi}\gamma_{(n)}\chi,\qquad\bar{\chi}^{\charge}\gamma_{(n)}\chi,\qquad\bar{\chi}\gamma_{9}\gamma_{(n)}\chi,\qquad\bar{\chi}^{\charge}\gamma_{9}\gamma_{(n)}\chi,
\end{equation}
where our notation is that these are $n$-forms
\begin{equation}
\bar{\chi}\gamma_{(n)}\chi=\frac{1}{n!}\bar{\chi}\gamma_{a_{1}\dots a_{n}}\chi\,e^{a_{1}\dots a_{n}},
\end{equation}
where $\{e^{a}\}$ is an orthonormal (co)frame for the unwarped metric on $X$. Some of the bilinears are real while others are complex depending on $n$. Our notation for the various bilinears is given in Table \ref{tab:bilinears}. Note that due the symmetries of bilinears of Majorana spinors in eight dimensions, the following quantities (and their duals) vanish identically:
\begin{equation}
\bar{\chi}^{\charge}\gamma_{9}\gamma_{(1)}\chi=0,\qquad\bar{\chi}^{\charge}\gamma_{9}\gamma_{(2)}\chi=0,\qquad\bar{\chi}^{\charge}\gamma_{(3)}\chi=0,\qquad\bar{\chi}^{\charge}\gamma_{(4)}\chi=0.
\end{equation}
Using the Fierz identities in \cite[Equation (1.3)]{1505.05238}, the scalar and one-form bilinears obey the following relations:\footnote{The bilinears of \cite{1505.05238} are related to ours as
\[
\begin{gathered}V_{+}=K,\qquad V_{-}=\re P,\qquad V_{3}=\im P,\qquad W=-L,\\
b_{+}=\zeta,\qquad b_{-}=\re S,\qquad b_{3}=\im S.
\end{gathered}
\]
}
\begin{equation}
\begin{aligned}\Vert\re P\Vert^{2}+(\re S)^{2} & =\Vert\im P\Vert^{2}+(\im S)^{2}, & L\lrcorner P & =\ii S.\\
\Vert K\Vert^{2}+\zeta^{2} & =1-\Vert\im P\Vert^{2}-(\im S)^{2},\qquad & L\lrcorner K & =0,\\
\Vert L\Vert^{2}+\Vert\im P\Vert^{2} & =1+(\re S)^{2}-\zeta^{2}, & K\lrcorner P & =-\zeta S,\\
\re P\lrcorner\im P+\re S\im S & =0.
\end{aligned}
\label{eq:fierz}
\end{equation}
These imply that the one-form $L$ satisfies
\begin{equation}
\Vert L\Vert^{2}=\Vert K\Vert^{2}+|S|^{2},\label{eq:Lsquared}
\end{equation}
and so $L$ vanishes if and only if both $K$ and $S$ vanish. 

From Table \ref{tab:bilinears} and the reality of spinor bilinears in Appendix \ref{sec:Gamma-matrices-in}, there are four one-form bilinears $(K,L,\re P,\im P)$. As we review in Section \ref{sec:Analysis-of--structures}, if they are linearly independent, $X$ is endowed with a local $\SU 2$ structure. On loci on $X$ where they are not linearly independent, the stabiliser of the Killing spinors enhances to $\SU 3$, $\Gx 2$ or $\SU 4$ depending on the rank of the kernel distribution of the four one-forms~\cite{1411.3493,1505.02270,1505.05238}.

\begin{table}
\noindent \begin{centering}
\begin{tabular}{ccccc}
\toprule 
$n$ & $\bar{\chi}\gamma_{(n)}\chi$ & $\bar{\chi}^{\charge}\gamma_{(n)}\chi$ & $\bar{\chi}\gamma_{9}\gamma_{(n)}\chi$ & $\bar{\chi}^{\charge}\gamma_{9}\gamma_{(n)}\chi$\tabularnewline
\midrule
\midrule 
0 & $1$ & $0$ & $\zeta$ & $S$\tabularnewline
1 & $K$ & $P$ & $\ii L$ & $0$\tabularnewline
2 & $\ii J$ & $0$ & $\ii\omega$ & $0$\tabularnewline
3 & $\ii\varphi$ & $0$ & $\phi$ & $\Omega$\tabularnewline
4 & $\Phi$ & $\Gamma$ & $\star\Phi$ & $\star\Gamma$\tabularnewline
5 & $-\star\phi$ & $-\star\Omega$ & $-\ii\star\varphi$ & $0$\tabularnewline
6 & $-\ii\star\omega$ & $0$ & $-\ii\star J$ & $0$\tabularnewline
7 & $\ii\star L$ & $0$ & $\star K$ & $\star P$\tabularnewline
8 & $\zeta\vol_{8}$ & $S\vol_{8}$ & $\vol_{8}$ & $0$\tabularnewline
\bottomrule
\end{tabular}
\par\end{centering}
\caption{Our definitions of spinor bilinears. Here $\protect\vol_{8}$ is the volume form defined by the unwarped metric on $X$.\label{tab:bilinears} }
\end{table}

\subsection{Differential and algebraic conditions\label{subsec:Differential-and-algebraic}}

In this subsection, we give the differential and algebraic conditions on the spinor bilinears of Table \ref{tab:bilinears}. These are computed from the Killing spinor equations given in Equation \eqref{eq:KS_eq}, and should be thought of as specifying the supergravity fluxes and the torsion of the underlying $G$-structure. Many of these expressions first appeared in some form in \cite{1212.6918,1311.5901}.\footnote{We found the Mathematica \cite{mathematica} packages \texttt{GammaMaP }\cite{1905.00429} and \texttt{Atlas2} \cite{atlas2} extremely useful for deriving these results. A pair of Mathematica notebooks with these calculations is included with the submission.}

\paragraph{Scalar conditions}

\begin{align}
\ee^{-3\Delta}\dd(\ee^{3\Delta}\zeta) & =f-4mK,\label{eq:zeta}\\
\ee^{-3\Delta}\dd(\ee^{3\Delta}S) & =-4mP.\label{eq:S}
\end{align}

\paragraph{One-form conditions}

\begin{align}
\dd(\ee^{3\Delta}K) & =0,\label{eq:K}\\
\dd(\ee^{3\Delta}P) & =0,\label{eq:P}\\
\ee^{-3\Delta}\dd(\ee^{3\Delta}L) & =-2mJ-\tfrac{1}{2}\omega\lrcorner F+\tfrac{1}{2}J\lrcorner\star F,\label{eq:L}
\end{align}
where $\lrcorner$ is the usual contraction of a $p$-vector into a $q$-form, defined in Appendix \ref{app:conventions}.

\paragraph{Two-form conditions}

\begin{align}
\ee^{-6\Delta}\dd(\ee^{6\Delta}J) & =f\wedge\omega-L\lrcorner\star F,\\
\ee^{-3\Delta}\dd(\ee^{3\Delta}\omega) & =-L\lrcorner F.\label{eq:omega}
\end{align}

\paragraph{Three-form conditions}

\begin{align}
\ee^{-3\Delta}\dd(\ee^{3\Delta}\varphi) & =f\lrcorner\star\varphi+J\cdot F,\label{eq:varphi-1}\\
\ee^{-6\Delta}\dd(\ee^{6\Delta}\phi) & =-4m\Phi+\zeta\,F-\star F,\label{eq:phi}\\
\ee^{-6\Delta}\dd(\ee^{6\Delta}\Omega) & =-4m\Gamma+S\,F,\label{eq:Omega}
\end{align}
where $J\cdot F$ denotes the action of $J$ as an endomorphism on the four-form $F$, defined in Appendix \ref{app:conventions}.

\paragraph{Four-form conditions}

\begin{align}
\ee^{-6\Delta}\dd(\ee^{6\Delta}\Phi) & =-K\wedge F,\\
\ee^{-6\Delta}\dd(\ee^{6\Delta}\Gamma) & =-P\wedge F,\label{eq:Gamma}\\
\ee^{-9\Delta}\dd(\ee^{9\Delta}\star\Phi) & =8m\star\phi+\Phi\wedge f-K\wedge\star F,\\
\ee^{-9\Delta}\dd(\ee^{9\Delta}\star\Gamma) & =8m\star\Omega+\Gamma\wedge f-P\wedge\star F.
\end{align}

\paragraph{Five-form conditions}

\begin{align}
\ee^{-9\Delta}\dd(\ee^{9\Delta}\star\phi) & =0,\\
\ee^{-9\Delta}\dd(\ee^{9\Delta}\star\Omega) & =0,\\
\ee^{-6\Delta}\dd(\ee^{6\Delta}\star\varphi) & =-L\lrcorner\star f-F\wedge\omega.
\end{align}

\paragraph{Six-form conditions}

\begin{align}
\ee^{-12\Delta}\dd(\ee^{12\Delta}\star\omega) & =f\wedge\star J,\\
\ee^{-9\Delta}\dd(\ee^{9\Delta}\star J) & =4m\star L.
\end{align}

\paragraph{Seven-form conditions}

\begin{align}
\dd\star L & =0,\label{eq:starL}\\
\ee^{-12\Delta}\dd(\ee^{12\Delta}\star K) & =-8m\star\zeta,\\
\ee^{-12\Delta}\dd(\ee^{12\Delta}\star P) & =-8m\star S.
\end{align}

\paragraph{Algebraic conditions}

In addition to the differential conditions above, there are also a number of algebraic conditions implied by the algebraic Killing spinor equation \eqref{eq:KS_eq_alg}. These can be derived by contracting the second equation of \eqref{eq:KS_eq} with $\bar{\chi}\gamma_{(n)}$, $\bar{\chi}^{\charge}\gamma_{(n)}$, $\bar{\chi}\gamma_{9}\gamma_{(n)}$ or $\bar{\chi}^{\charge}\gamma_{9}\gamma_{(n)}$, and taking sums and differences of the resulting identities. Given a \emph{non-vanishing} chiral spinor $\xi_{+}$, the eight spinors $\gamma_{a}\xi^{+}$ form a basis of the space of negative-chirality spinors~\cite[Section 2.6]{Babalic:2014dea}, and vice verse for an antichiral spinor $\xi^{-}$. If the Killing spinors $\chi_{i}$ could be written as a sum of non-vanishing chiral and antichiral spinors, it would be sufficient to project \eqref{eq:KS_eq_alg} using $\bar{\chi}\gamma_{(1)}$ and $\bar{\chi}^{\charge}\gamma_{(1)}$ alone, giving a number of one-form conditions. This is not the case for us, since the Killing spinors can become chiral or antichiral over certain loci of $X$. Instead, generically one needs both the zero-form and one-form relations~\cite{1311.5901}.\footnote{If the local structure group is enhanced, some of the zero- or one-form conditions trivialise, and one needs to impose some of the higher-rank conditions. This happens, for example, in the case of a strict $\SU 4$ structure.} For the reader's convenience, we give the scalar and one-form conditions below, and include up to the four-form conditions in Appendix \ref{sec:Algebraic-Killing-spinor} -- the remaining identities are simply the Hodge dual of these.\footnote{These expressions appeared in a slightly different form in \cite{1212.6918,1311.5901}.}
\begin{align}
0 & =-2m\zeta+K\lrcorner\dd\Delta-\tfrac{1}{6}\Phi\lrcorner F, & 0 & =\dd\Delta-\tfrac{1}{3}\zeta f+\tfrac{1}{6}F\lrcorner(\star\phi),\label{eq:alg1}\\
0 & =L\lrcorner f, & 0 & =2mL-\dd\Delta\lrcorner J+\tfrac{1}{3}f\lrcorner\omega+\tfrac{1}{6}\varphi\lrcorner F,\label{eq:f_cond}\\
0 & =-2mS+P\lrcorner\dd\Delta-\tfrac{1}{6}\Gamma\lrcorner F, & 0 & =-\tfrac{1}{3}Sf+\tfrac{1}{6}F\lrcorner(\star\Omega),\label{eq:S_nonzero}\\
0 & =-2m+\tfrac{1}{3}K\lrcorner f-\tfrac{1}{6}(\star\Phi)\lrcorner F, & 0 & =-\dd\Delta\lrcorner\omega+\tfrac{1}{3}f\lrcorner J+\tfrac{1}{6}F\lrcorner(\star\varphi),\\
0 & =L\lrcorner\dd\Delta, & 0 & =2mK+\zeta\,\dd\Delta-\tfrac{1}{3}f+\tfrac{1}{6}\phi\lrcorner F,\label{eq:L_cond}\\
0 & =\tfrac{1}{3}P\lrcorner f-\tfrac{1}{6}(\star\Gamma)\lrcorner F, & 0 & =2mP+S\,\dd\Delta+\tfrac{1}{6}\Omega\lrcorner F.\label{eq:P.f}
\end{align}

\subsection{Bianchi identities and equations of motion\label{subsec:Bianchi-identities}}

The eleven-dimensional Bianchi identity and equation of motion for the four-form $G$ are equivalent to the reduced conditions for $F$ and $f$ given in \eqref{eq:Bianchi (1,0)} and \eqref{eq:implied}. For the $N=(1,0)$ case discussed in \cite{Martelli:2003ki,hep-th/0511047}, the supersymmetry conditions together with \eqref{eq:Bianchi (1,0)} imply that all the remaining equations of motion and Bianchi identities are satisfied. Since there is always an $(1,0)$ subsector present in the backgrounds we consider, we again know that, at most, we will have to impose the two equations in \eqref{eq:Bianchi (1,0)} in addition to supersymmetry. We now show that when the background is sufficiently generic, both of the equations in \eqref{eq:Bianchi (1,0)} are actually implied by the Killing spinor equations, and so they do not have to be imposed separately, i.e.~an $N=(2,0)$ solution to the Killing spinor equations \eqref{eq:KS_eq} and \eqref{eq:KS_eq_alg} automatically solves all of the supergravity equations of motion and Bianchi identities.

Let us first see how \eqref{eq:implied} are implied by $N=(1,0)$ supersymmetry and \eqref{eq:Bianchi (1,0)}. We focus on the differential conditions that survive in the $(1,0)$ limit. Taking an exterior derivative of the differential condition for $\zeta$ in \eqref{eq:zeta} gives
\begin{equation}
\dd(\ee^{3\Delta}f)=4m\,\dd(\ee^{3\Delta}K)=0,\label{eq:bianchi_f}
\end{equation}
where in the second equality we used $\dd(\ee^{3\Delta}K)=0$ from \eqref{eq:K}. Thus, the Bianchi identity for $f$ is implied by supersymmetry. Next, taking an exterior derivative of the differential condition for $\phi$ in \eqref{eq:phi} gives
\begin{equation}
\begin{split}0 & =-4m\,\dd(\ee^{6\Delta}\Phi)+\dd(\ee^{3\Delta}\zeta)\wedge\ee^{3\Delta}F+\ee^{3\Delta}\zeta\,\dd(\ee^{3\Delta}F)-\dd(\ee^{6\Delta}\star F)\\
 & =\ee^{3\Delta}\zeta\,\dd(\ee^{3\Delta}F)-\left(\dd(\ee^{6\Delta}\star F)-\ee^{6\Delta}f\wedge F\right).
\end{split}
\end{equation}
Assuming the Bianchi identity for $F$ in \eqref{eq:Bianchi (1,0)} is satisfied, supersymmetry implies the equation of motion for $F$. Thus, $(1,0)$ supersymmetry plus \eqref{eq:Bianchi (1,0)} implies the equations of motion and Bianchi identity for $G$ (with the remaining equations of motion following from the argument of \cite{Martelli:2003ki}).

Given that we have $(2,0)$ supersymmetry, we might hope that \eqref{eq:Bianchi (1,0)} are themselves implied by supersymmetry. Since these conditions are not implied by the $(1,0)$ equations, we should look at the extra equations that come from extended supersymmetry. Taking the exterior derivative of the differential condition \eqref{eq:Omega} for $\Omega$ gives
\begin{equation}
\begin{split}0 & =-4m\,\dd(\ee^{6\Delta}\Gamma)+\dd(\ee^{3\Delta}S)\wedge\ee^{3\Delta}F+\ee^{3\Delta}S\,\dd(\ee^{3\Delta}F)\\
 & =\ee^{3\Delta}S\,\dd(\ee^{3\Delta}F),
\end{split}
\end{equation}
where we used the derivatives of $S$ and $\Gamma$ from \eqref{eq:S} and \eqref{eq:Gamma}. For $S\neq0$, $N=(2,0)$ supersymmetry thus implies the Bianchi identity for $F$. Next, consider the Hodge star of the one-form condition \eqref{eq:S_nonzero} derived from the algebraic KS equation:
\begin{equation}
0=-\tfrac{1}{3}S\star f-\tfrac{1}{6}F\wedge\Omega.
\end{equation}
Multiplying by $\ee^{9\Delta}$ and taking a derivative of this equation gives
\begin{equation}
\begin{split}0 & =-2\dd(\ee^{3\Delta}S)\wedge\ee^{6\Delta}\star f-2\ee^{3\Delta}S\,\dd(\ee^{6\Delta}\star f)-\dd(\ee^{3\Delta}F)\wedge\ee^{6\Delta}\Omega-\ee^{3\Delta}F\wedge\dd(\ee^{6\Delta}\Omega)\\
 & =8m\ee^{9\Delta}P\wedge\star f+4m\ee^{9\Delta}F\wedge\Gamma-2\ee^{3\Delta}S\,\dd(\ee^{6\Delta}\star f)-S\,\ee^{9\Delta}F\wedge F,
\end{split}
\end{equation}
where we used the derivatives of $S$ and $\Omega$ and from \eqref{eq:S} and \eqref{eq:Omega}, and have assumed $S\neq0$ so that the Bianchi identity for $F$ is automatically satisfied. Finally, we note that the dual of the zero-form relation \eqref{eq:P.f} reads
\begin{equation}
0=2P\wedge\star f+F\wedge\Gamma,
\end{equation}
and so inserting this above gives
\begin{equation}
0=-2\ee^{3\Delta}S\left(\dd(\ee^{6\Delta}\star f)+\tfrac{1}{2}\ee^{6\Delta}F\wedge F\right).
\end{equation}
Thus, if $S\neq0$, the equation of motion for $f$ is also implied by supersymmetry.

In conclusion, if $S\neq0$, an $N=(2,0)$ solution to the Killing spinor equations \eqref{eq:KS_eq} and \eqref{eq:KS_eq_alg} automatically satisfies the equations of motion and Bianchi identities for both $F$ and $f$. The results of \cite{Martelli:2003ki} then imply that all of the remaining supergravity equations of motion are also satisfied. If $S$ vanishes, one needs to check that \eqref{eq:Bianchi (1,0)} hold, which often leads to extra constraints on the structure.

\subsection{A Killing vector\label{subsec:A-Killing-vector}}

The geometries we have focused on preserve $(2,0)$ supersymmetry in the boundary conformal field theory. The boundary CFT thus admits a symmetry which rotates the supercharges by $\SO 2\simeq\Uni 1$ -- such a symmetry is known as an R-symmetry. We expect this to be present in the dual AdS$_{3}$ geometry through the existence of a nowhere-vanishing Killing vector which preserves the full supergravity solution but rotates the complex Killing spinor by a phase. We will refer to this Killing vector as the \emph{R-symmetry vector}.

It is straightforward to show that the Killing spinor equations \eqref{eq:KS_eq} and \eqref{eq:KS_eq_alg} imply that the one-form $L$ satisfies\footnote{Note that, leaving the parameter $m_{2}$ unfixed, one finds that $\nabla_{(a}L_{b)}$ is proportional to $(m-m_{2})$ times an expression that is generically non-zero. Without fine-tuning so that this expression vanishes, $L$ is not a Killing vector for $m_{2}=-m$, and so $N=(1,1)$ backgrounds do not admit an R-symmetry Killing vector.}
\begin{equation}
\nabla_{(a}L_{b)}=0,
\end{equation}
and so the vector field dual to $L$ is Killing, $\mathcal{L}_{L}g=0$.\footnote{This also agrees with the differential condition \eqref{eq:starL} derived from the Killing spinor equations, since one has the identity $\dd\star L=-\mathcal{L}_{L}\vol_{8}$, which must vanish if $L$ is Killing.} As we discuss in Section \ref{subsec:L=00003D0}, if the norm of $L$ vanishes somewhere on $X$, one has $m=0$ there. However, since $m$ is assumed to be \emph{constant} on $X$, such backgrounds are necessarily Minkowski. Conversely, if the background is AdS, so that $m\neq0$, $L$ is nowhere-vanishing on $X$. Thus, AdS$_{3}$ backgrounds with $(2,0)$ supersymmetry necessarily have a nowhere-vanishing Killing vector, given by the dual of $L$.

At this point, we have shown only that $L$ is an isometry of the metric. For $L$ to preserve the full solution, and so be identified as a candidate R-symmetry vector field, we need to show that it preserves the remaining supergravity fields $(\Delta,f,F)$. The zero-form conditions in \eqref{eq:f_cond} and \eqref{eq:L_cond} imply that both $\dd\Delta$ and $f$ have no components along $L$, so
\begin{align}
\begin{aligned}\mathcal{L}_{L}\Delta & =\imath_{L}\dd\Delta=0,\\
\mathcal{L}_{L}(\ee^{3\Delta}f) & =\imath_{L}\dd(\ee^{3\Delta}f)+\dd(\ee^{3\Delta}\imath_{L}f)=0,
\end{aligned}
\end{align}
where in the second line we used the Bianchi identity for $f$, which, as we saw in the previous subsection, is implied by supersymmetry alone. We also have
\begin{equation}
\mathcal{L}_{L}(\ee^{3\Delta}F)=\dd\imath_{L}(\ee^{3\Delta}F)+\imath_{L}\dd(\ee^{3\Delta}F)=\imath_{L}\dd(\ee^{3\Delta}F),
\end{equation}
where we used that the differential condition for $\omega$ in \eqref{eq:omega} implies $\imath_{L}(\ee^{3\Delta}F)$ is $\dd$-closed. Putting this together, we see that $L$ preserves the full supergravity solution if the Bianchi identity for $F$ holds. From the discussion in the previous subsection, with $S\neq0$, $(2,0)$ supersymmetry is sufficient for this Bianchi identity to hold, and so $L$ will automatically preserve the full solution. For backgrounds with $S=0$, one needs to impose the Bianchi identity separately. Either way, the vector field dual to $L$ is a nowhere-vanishing Killing vector which preserves the full solution.

A final point to check is how this $\Uni 1$ acts on the Killing spinors themselves. Since the R-symmetry should rotate the external spinors $\psi_{i}$ and the eleven-dimensional spinor $\eta$ is invariant, the Killing spinors should transform under the $\Uni 1$ -- indeed one can show that the action of $L$ on the complex spinor $\chi$ is
\begin{equation}
\mathcal{L}_{L}\chi=-2\ii m\chi,
\end{equation}
and so $\chi$ is charge $-2m$ under the $\Uni 1$. The easiest way to see this is in the AdS case with $S\neq0$. Following the argument in \cite{1207.3082}, since $L$ preserves the bosonic fields, the Lie derivatives of the scalar bilinears can be used to fix the R-charge of the Killing spinor. The differential condition on $S=\bar{\chi}^{\charge}\gamma_{9}\chi$ in \eqref{eq:S} implies
\begin{equation}
\mathcal{L}_{L}S=\imath_{L}\dd S=-4m\imath_{L}P=-4\ii mS,\label{eq:S_deriv}
\end{equation}
where we used $\imath_{L}P=\ii S$ from \eqref{eq:fierz}, and so $\chi$ is charge $-2m$. Putting this all together, we see that $L$ is indeed a nowhere-vanishing Killing vector which preserves the full supergravity solution but rotates the complex Killing spinor by a phase, and so should be identified as the vector field generating the $\Uni 1$ R-symmetry of the boundary $(2,0)$ superconformal field theory.

Since $L$ preserves the metric (and hence the gamma matrices), the charges of the remaining bilinears can be read off from whether they are constructed as $\bar{\chi}^{\charge}\dots\chi$ or $\bar{\chi}\dots\chi$, with the former charge $-4m$ and the latter uncharged. In particular, only $S$, $P$, $\Omega$ and $\Gamma$ of Table \ref{tab:bilinears} are charged under the $\Uni 1$ generated by $L$. One can also compute symmetric derivatives of the real vector $K$ and the complex vector $P$ to check for further Killing vectors. These are given by
\begin{equation}
\nabla_{(a}K_{b)}=-2m\zeta g_{ab}-\tfrac{1}{12}\Phi^{efcd}F_{efca}g_{db},\qquad\nabla_{(a}P_{b)}=-2mSg_{ab}-\tfrac{1}{12}\Gamma^{efcd}F_{efca}g_{db}.\label{eq:K_P_Killing}
\end{equation}
Generally, neither of these expressions vanishes, so there are no further Killing vectors. There are, however, special cases where $K$ or $P$ can also be Killing.

\section{Local \texorpdfstring{$G$}{G}-structures and stabiliser groups\label{sec:Analysis-of--structures}}

The analysis we have presented so far has been completely general in the sense that we have assumed only the existence of two nowhere-vanishing Majorana spinors. The question of the local $G$-structure defined by such a pair on an eight-manifold is surprisingly subtle. Fortunately, this problem has been tackled in great detail in a series of papers by Babalic and Lazaroiu~\cite{1411.3493,1505.02270,1505.05238}. Their analysis can be summarised as follows. A choice of two nowhere-vanishing Majorana spinors $(\chi_{1},\chi_{2})$ on an eight-manifold generically defines four one-forms -- in our notation, these are $(L,K,\re P,\im P)$. The Fierz identities \eqref{eq:fierz} obeyed by these one-forms do not force them to be linearly dependent in general, and so generically one has four linearly independent one-forms which define a rank-four distribution on $X$. The space transverse to the distribution is four-dimensional and admits an $\SU 2$ structure. On certain loci of $X$, one finds that some of the one-forms can become linearly dependent, and the distribution reduces in rank to two, one or zero, corresponding to cases where the transverse space admits an $\SU 3$, $\Gx 2$ or $\SU 4$ structure. In what follows, we summarise the main results of \cite{1505.02270} in our notation and discuss how the local $G$-structure is classified by the values of the scalar bilinears and the norm of the one-forms.

\begin{figure}
\includegraphics[viewport=0bp 35.9129bp 198bp 167.5937bp,clip]{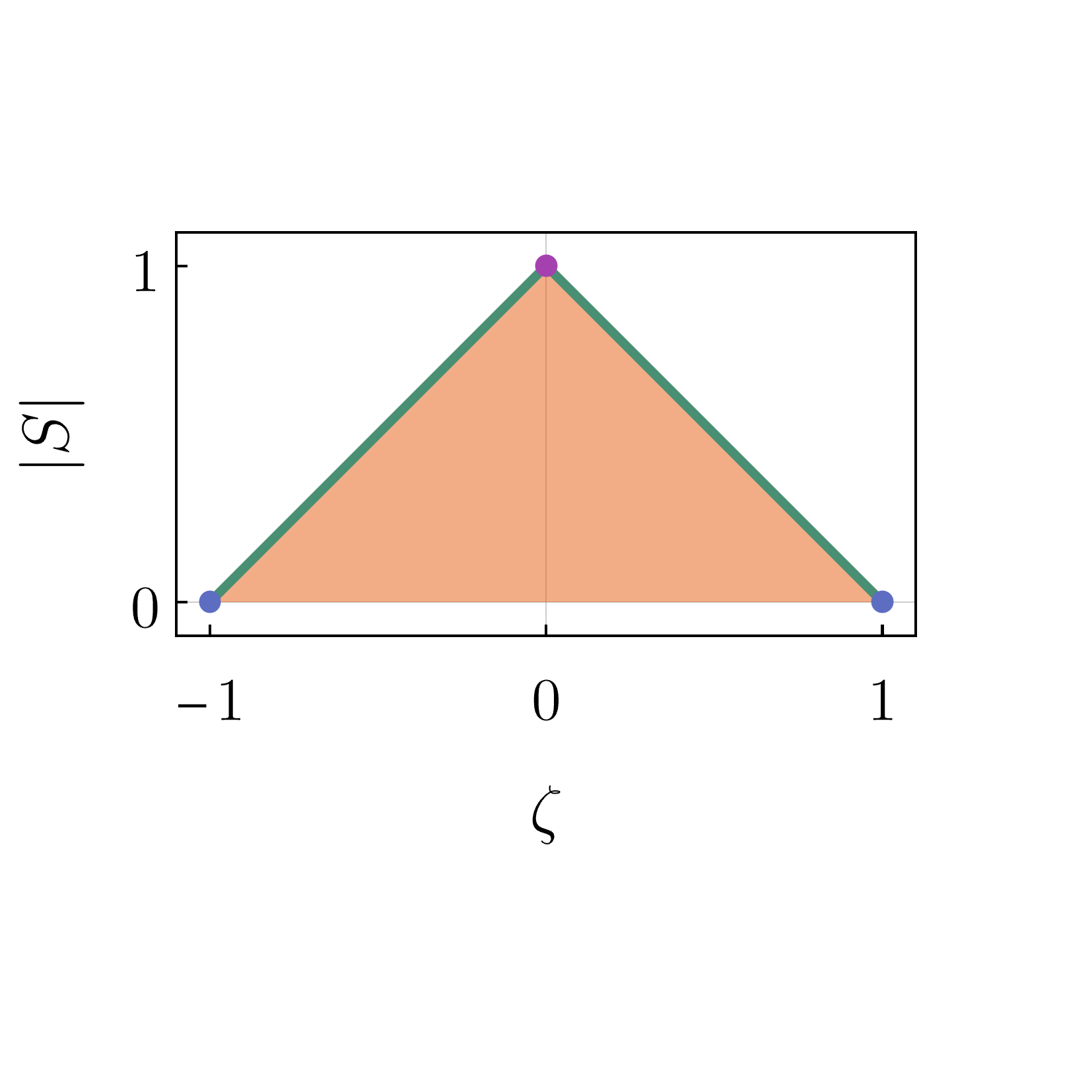}\caption{A illustration of the space $\mathcal{R}$ on the $(\zeta,|S|)$-plane. The three-dimensional space $\mathcal{R}$ can be obtained by revolving this around the $\zeta$-axis by $2\pi$. The blue points at $(\pm1,0)$ correspond to $\protect\SU 4$ structures. The purple point at $(0,1)$ is a $\protect\Gx 2$ structure. The green lines correspond to $\protect\SU 3$ structures. The stabiliser of spinors that map to the orange interior can be $\protect\SU 2$ or $\protect\SU 3$.\label{fig:A-illustration-of}}

\end{figure}

The key idea is that the Fierz identities in \eqref{eq:fierz} constrain the allowed values of the scalars and the norms of the one-forms. As is shown in \cite[Section 3.2]{1505.02270}, given the definitions of $\zeta$ and $S$ as spinor bilinears, the Cauchy--Schwarz inequality and the Fierz identities constrain the scalars to take values in a certain subset $\mathcal{R}\subset\bR^{3}$, defined as
\begin{equation}
\mathcal{R}=\bigl\{(\zeta,\re S,\im S)\in[-1,1]\;\big|\;|S|+|\zeta|\leq1\bigr\}.
\end{equation}
Given that $|S|$ lies in $[0,1]$, $\mathcal{R}$ is the set of points in the $(\re S,\im S)$ plane that satisfy $|S|\leq1-|\zeta|$ rotated around the $\zeta$-axis, so that $\mathcal{R}$ is the union of two right-angled cones with a common base. An illustration of $\mathcal{R}$ on the $(\zeta,|S|)$-plane is given in Figure \ref{fig:A-illustration-of}. Equivalently, $\mathcal{R}$ can be described by polynomial inequalities as
\begin{equation}
\mathcal{R}=\bigl\{(\zeta,\re S,\im S)\;\big|\;\zeta^{2}+|S|^{2}\leq1,\,|S|^{2}\leq\tfrac{1}{4}(1+|S|^{2}-\zeta^{2})^{2}\bigr\},
\end{equation}
where the first condition gives \eqref{eq:scalar_norm}. The boundary of $\mathcal{R}$, denoted by $\partial\mathcal{R}$, is the subset
\begin{equation}
\partial\mathcal{R}=\bigl\{(\zeta,\re S,\im S)\in[-1,1]\;\big|\;|S|+|\zeta|=1|\bigr\}.
\end{equation}
As discussed in \cite{1505.02270}, there is a canonical stratification of both $\mathcal{R}$ and $\partial\mathcal{R}$ into sets of points which satisfy polynomial inequalities (``semi-algebraic sets'').

The interval
\begin{equation}
I=\bigl\{(\zeta,0,0)\;\big|\;\zeta\in[-1,1]\bigr\}
\end{equation}
lies along the $\zeta$-axis with boundary $\partial I=\{\zeta=\pm1\}$, while the disk
\begin{equation}
D=\bigl\{(0,\re S,\im S)\;\big|\;|S|^{2}\leq1\bigr\}
\end{equation}
is the base shared by the right-angled cones. The boundary $\partial D$ of the disk is a circle such that
\begin{equation}
\partial D=\bigl\{(0,\re S,\im S)\;\big|\;|S|^{2}=1\bigr\}.
\end{equation}
$\mathcal{R}$ itself is a disjoint union of its interior, $\op{int}\mathcal{R}$, and its boundary
\begin{equation}
\mathcal{R}=\op{int}\mathcal{R}\sqcup\partial\mathcal{R}.
\end{equation}
The boundary decomposes as a disjoint union of the two tips of the cones $\partial_{0}\mathcal{R}=\partial I$, the circle bounding the disk $\partial_{1}\mathcal{R}=\partial D$, and the surface of the two cones $\partial_{2}\mathcal{R}$ (and are of dimension zero, one, and two respectively):
\begin{equation}
\partial\mathcal{R}=\partial I\sqcup\partial D\sqcup\partial_{2}\mathcal{R}.
\end{equation}

The eight-manifold $X$ can be partitioned into subspaces according to whether the image of a point $p\in X$ under the map $(\zeta,\re S,\im S)\colon X\to\mathcal{R}$ lies on the boundary, $\partial\mathcal{R}$, or in the interior, $\op{int}\mathcal{R}$. We denote the subspace whose points are mapped to the boundary by $\mathcal{S}\subset X$, with its complement in $X$ given by $\mathcal{G}=X\setminus\mathcal{S}$:
\begin{equation}
X=\mathcal{S}\sqcup\mathcal{G}.
\end{equation}
Defining $\mathcal{S}_{ij}$ to be the locus on $X$ where $(\chi_{1},\chi_{2})$ can be written in terms of $i$ antichiral and $j$ chiral spinors,\footnote{The somewhat unorthodox labelling matches that used in \cite{1505.02270}. } there is a further stratification as
\begin{equation}
\begin{gathered}\im\mathcal{S}_{11}=\partial D,\\
\im\mathcal{S}_{12}=\partial_{2}\mathcal{R}^{+},\qquad\im\mathcal{S}_{21}=\partial_{2}\mathcal{R}^{-},\\
\im\mathcal{S}_{02}=\partial I^{+},\qquad\im\mathcal{S}_{20}=\partial I^{-},\\
\im\mathcal{G}=\im\mathcal{S}_{22}=\op{int}\mathcal{R},
\end{gathered}
\end{equation}
where, for example, $\partial_{2}\mathcal{R}^{+}$ are points in $\partial_{2}\mathcal{R}$ with $\zeta>0$, and likewise for the other subsets. The set $\mathcal{S}$ is special in that one or both Majorana spinors become chiral or antichiral, whereas the locus $\mathcal{G}$ should be thought of as the ``generic'' case where neither spinor has definite chirality (and so must be written using two chiral and two antichiral spinors). The stabiliser $H_{p}$ of the pair of Majorana spinors at a point $p$ is then related to the image of the point in $\mathcal{R}$ as
\begin{equation}
\begin{aligned}p\in\mathcal{S}_{02}\cup\mathcal{S}_{20}\colon & H_{p}=\SU 4, & \qquad p\in\mathcal{S}_{12}\cup\mathcal{S}_{21}\colon & H_{p}=\SU 3,\\
p\in\mathcal{S}_{11}\colon & H_{p}=\Gx 2, & p\in\mathcal{G}=\mathcal{S}_{22}\colon & H_{p}=\SU 2\text{ or }\text{\ensuremath{\SU 3}}.
\end{aligned}
\end{equation}
The chirality of the spinors is sufficient to differentiate between $\SU 4$, $\Gx 2$ and $\SU 3$ structures, whereas if $p\in\mathcal{G}$ the chirality alone is not enough to tell us whether there is an $\SU 2$ or $\SU 3$ structure. Instead, the stabiliser is fixed by the rank of the intersection of the kernel distributions of the four one-forms $(K,L,\re P,\im P)$ -- if the resulting distribution is rank four, it admits an $\SU 2$ structure, whereas if it is rank six, seven or eight it carries an $\SU 3$, $\Gx 2$ or $\SU 4$ structure respectively. Generically the distribution defined by the kernel of the four one-forms is not integrable. However, \eqref{eq:K} and \eqref{eq:P} imply that the distribution defined by the kernel of $(K,\re P,\im P)$ is integrable, giving a (singular) foliation of $X$ where the five-dimensional transverse space admits an $\SU 2$ structure~\cite{1505.02270}.

We note in passing that for $\AdS 3$ compactifications, the number of supersymmetries preserved by a spacetime-filling M2-brane placed at a point $p\in X$ is given by the number of purely chiral solutions to the Killing spinor equations~\cite{1311.5901}. For example, for $p\in\mathcal{S}_{02}$, an M2-brane would preserve $(2,0)$ supersymmetry. An M2-brane placed on $\mathcal{S}_{11}$ or $\mathcal{S}_{12}$ would preserve $(1,0)$ supersymmetry, while $\mathcal{S}_{22}$ preserves no supersymmetry.

As we mentioned above, the stabiliser of $(\chi_{1},\chi_{2})$ is fixed by the rank of the intersection of the kernel distributions of the four one-forms. Equivalently, working at a point on $X$, we can ask about the dimension of the vector space spanned by the one-forms; for example, at points where the one-forms $(K,L,\re P,\im P)$ are linearly independent, the Killing spinors are stabilised by $\SU 2$. If the one-forms span a two-, one- or zero-dimensional vector space, the Killing spinors are stabilised by $\SU 3$, $\Gx 2$ or $\SU 4$ respectively. We want to understand the constraints which lead to these various cases. First, one defines the function $\beta\colon X\to\bR$ as
\begin{equation}
\beta\equiv\sqrt{(\im S)^{2}+\Vert\im P\Vert^{2}}=\sqrt{(\re S)^{2}+\Vert\re P\Vert^{2}},\label{eq:beta}
\end{equation}
where the second equality follows from the Fierz identities. As proved in \cite{1505.05238}, the local stabiliser can be specified by the value of $(\zeta,\re S,\im S,\beta)$, which, compared to the data contained in the image of a point in $\mathcal{R}$, amounts to information about the norm of $\re P$ or $\im P$. 

In more detail, recall that given a set of vectors $\{v_{i}\}$, one can form their Gram matrix as $G_{ij}=v_{i}\cdot v_{j}$, with the dimension of the space spanned by the vectors equal to the rank of the Gram matrix. The Fierz identities \eqref{eq:fierz} fix the Gram matrix of $(K,L,\re P,\im P)$ in terms of the map $(\zeta,\re S,\im S,\beta)$, and so the stabiliser of the Killing spinors can be studied by understanding how the Gram matrix changes in rank. As shown in \cite{1505.02270}, $(\zeta,\re S,\im S,\beta)$ takes values in a certain four-dimensional space $\mathfrak{B}$:
\begin{equation}
\mathfrak{B}=\bigl\{(\zeta,\re S,\im S,\beta)\in\mathbb{R}^{4}\;\big|\;(\zeta,\re S,\im S)\in\mathcal{R},\,\beta\in\mathcal{J}(\zeta,S)\bigr\},
\end{equation}
where $\mathcal{J}(\zeta,S)$ is the closed interval
\begin{equation}
\mathcal{J}(\zeta,S)=\Bigl[\sqrt{f_{-}(\zeta,S)},\sqrt{f_{+}(\zeta,S)}\Bigr],
\end{equation}
and $f_{\pm}\colon\mathcal{R}\to\mathbb{R}$ are the roots of the polynomial
\begin{equation}
f^{2}-(1+|S|^{2}-\zeta^{2})f+|S|^{2}.\label{eq:fpm}
\end{equation}
Note that the roots satisfy
\begin{equation}
0\leq|S|^{2}\leq f_{-}(\zeta,S)\leq f_{+}(\zeta,S)\leq1,
\end{equation}
with $f_{-}=0$ for $(0,\re S,\im S)\in I$, $f_{+}=1$ for $(0,\re S,\im S)\in D$, and $f_{-}=f_{+}\equiv|S|$ for $(0,\re S,\im S)\in\partial\mathcal{R}$. The space $\mathfrak{B}$ can be thought of as fibred over $\mathcal{R}$ via the projection map $\pi\colon(\zeta,\re S,\im S,\beta)\to(\zeta,\re S,\im S)$. The fibre over $(\zeta,\re S,\im S)\in\mathcal{R}$ is the interval $\mathcal{J}(\zeta,S)$ which degenerates to a single point $\{|S|\}$ over $\partial\mathcal{R}$. The classification of the stabiliser also depends on the following function
\begin{equation}
g(S,\beta)=\frac{1}{\beta}\sqrt{(1-\beta^{2})(\beta^{2}-|S|^{2})}.\label{eq:g}
\end{equation}

\begin{figure}
\includegraphics[viewport=0bp 0bp 197bp 149bp,clip]{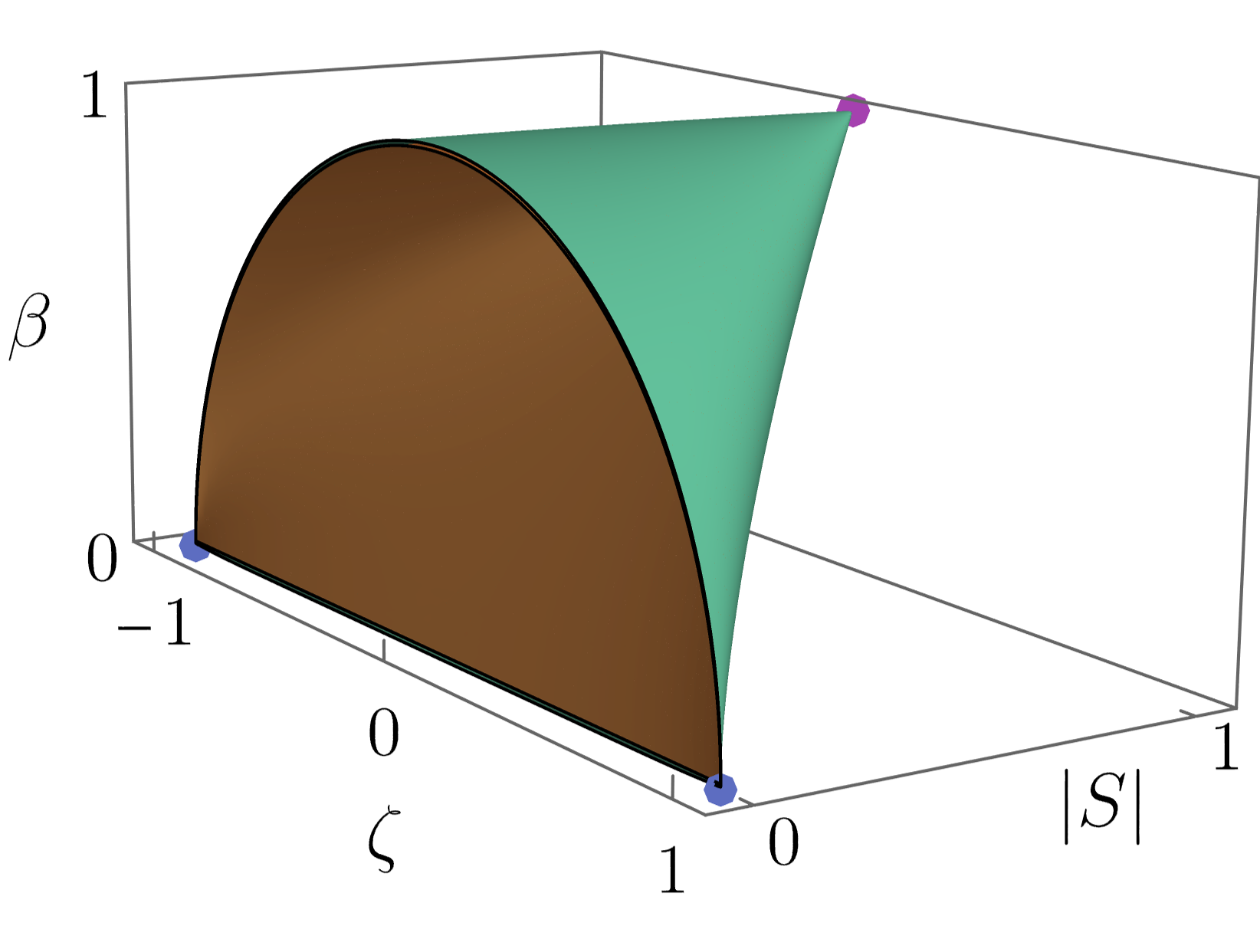}\quad{}\includegraphics[viewport=0bp 23.94196bp 198bp 167.5937bp,clip]{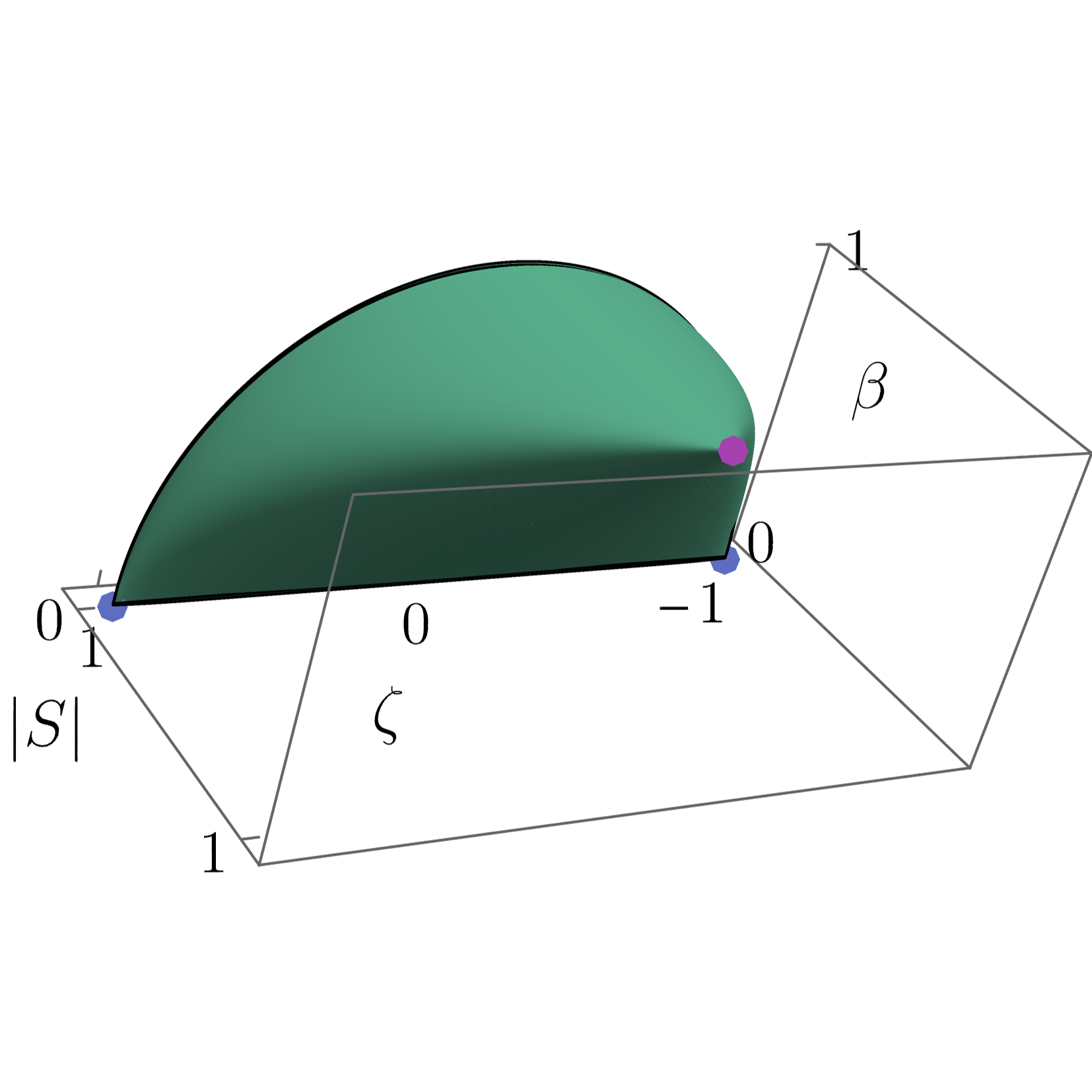}\\
\includegraphics[viewport=0bp 35.9129bp 198bp 167.5937bp,clip]{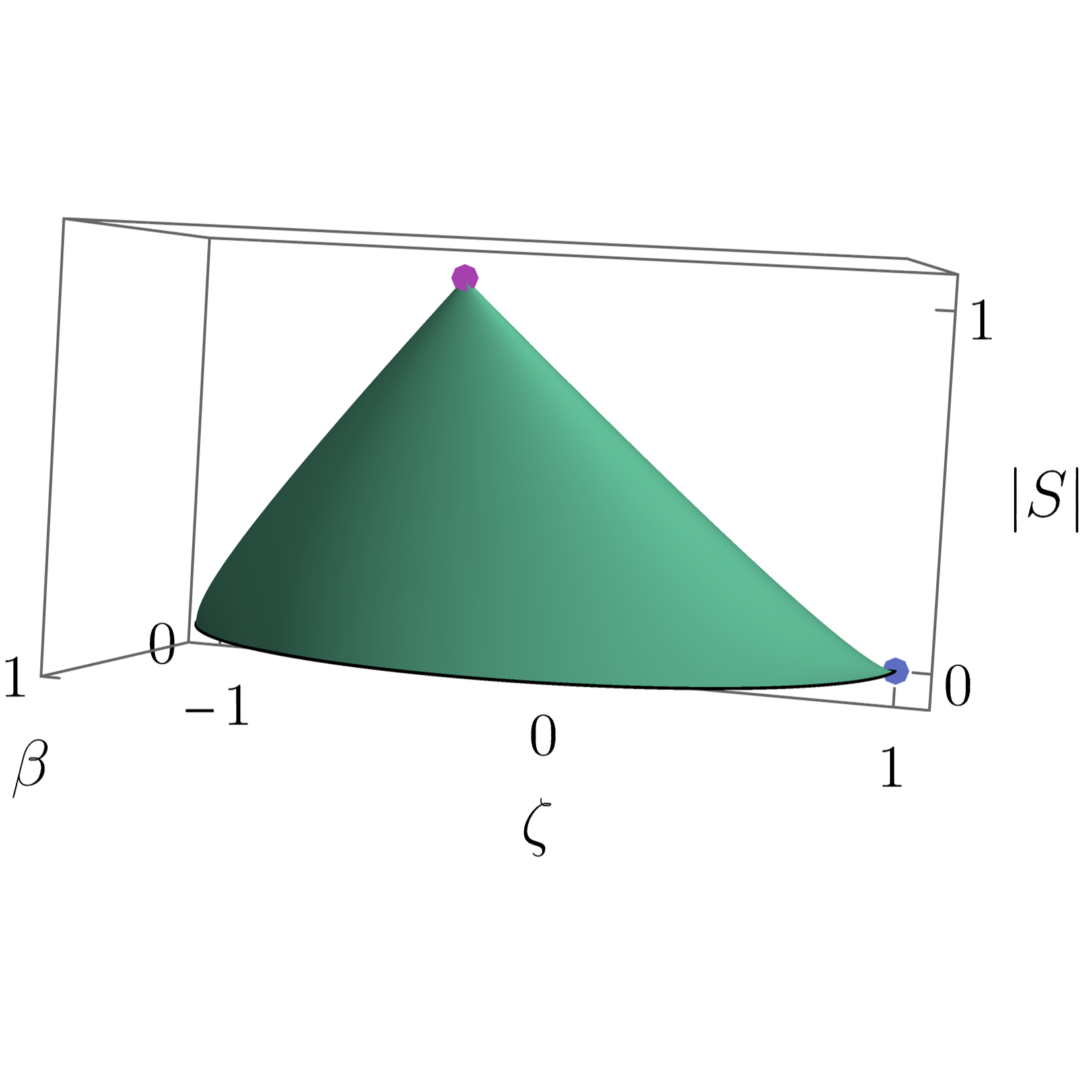}\quad{}\caption{Illustrations of the allowed stabiliser groups in $(\zeta,|S|,\beta)$ space. The colour of the regions match those in Figure \ref{fig:A-illustration-of}. The blue points at $(\pm1,0,0)$ correspond to $\protect\SU 4$ structures. The purple point at $(0,1,1)$ is a $\protect\Gx 2$ structure. The green surface that wraps around (though does not fully enclose) the space corresponds to $\protect\SU 3$ structures. The orange interior corresponds to spinors stabilised by an $\protect\SU 2$.\label{fig:Illustrations-of-the}}
\end{figure}

The results of \cite{1505.02270} are summarised in Table \ref{tab:Stabiliser-group-of}. We give an illustration of these results in $(\zeta,|S|,\beta)$ space in Figure \ref{fig:Illustrations-of-the}. One sees that the admissible values of $(\zeta,|S|,\beta)$ fill out a compact three-dimensional space. The local structure defined by the spinors is generically $\SU 2$, with the structure enhancing to $\SU 3$ along certain surfaces, and to $\Gx 2$ or $\SU 4$ at isolated points.
\begin{table}
\noindent \begin{centering}
\begin{tabular}{cccccc}
\toprule 
$(\zeta,\re S,\im S)$-image & $\zeta$ & $|S|$ & $\beta$ & $H_{p}$ & Example\tabularnewline
\midrule
\midrule 
$\partial I^{\pm}$ & $\pm1$ & $0$ & $0$ & $\SU 4$ & \S\ref{subsec:Strict--su4}\tabularnewline
$\partial D$ & $0$ & $1$ & $1$ & $\Gx 2$ & \S\ref{subsec:-holonomy-solutions}\tabularnewline
$\op{int}I$ & $(-1,1)$ & $0$ & $0$ & $\SU 3$ & \S\ref{subsec:All-purely-electric}, \S\ref{subsec:AdS-from-wrapping}\tabularnewline
$\op{int}D$ & $0$ & $[0,1)$ & $1$ & $\SU 3$ & \S\ref{subsec:AdS-from-wrapping-1}\tabularnewline
$\op{int}D\backslash\{0\}$ & $0$ & $(0,1)$ & $|S|$ & $\SU 3$ & \S\ref{subsec:New-example}\tabularnewline
$\op{int}\mathcal{R}^{\pm}$ & $\pm g(S,\beta)$ & $[0,1)$ & $(|S|,1)$ & $\SU 3$ & \S\ref{subsec:SLAG-three-cycles-in}, \S\ref{subsec:Generic-case-with}\tabularnewline
$\op{int}\mathcal{R}$ & $(-1,1)$ & $[0,1)$ & $\mathcal{J}(\zeta,|S|)$ & $\SU 2$ & \S\ref{sec:Examples-with-su2}\tabularnewline
\bottomrule
\end{tabular}
\par\end{centering}
\caption{Stabilisers of $(\chi_{1},\chi_{2})$ at a point $p\in X$ for various choices of $\xi$, $|S|$ and $\beta$. The first column gives the image of $(\zeta,\protect\re S,\protect\im S)$ in $\mathcal{R}$, the remaining columns give the values of $\zeta$, $|S|$ and $\beta$, the stabiliser group, and the section of the paper where we give an example of the class of background. \label{tab:Stabiliser-group-of}}
\end{table}

\section{Examples\label{sec:Examples}}

With a description of the various stabilisers in hand, we now give examples of known AdS$_{3}$ backgrounds in the literature which are described by the differential and algebraic conditions derived from the Killing spinor equations, including Calabi--Yau fourfolds, manifolds of $\Gx 2$ holonomy, and AdS solutions that come from wrapping M5-branes on Kähler or co-associative four-cycles. We also generalise the Minkowski solutions found in \cite[Section 5]{Martelli:2003ki} describing M5-branes wrapping SLAG three-cycles in eight-manifolds with an $\SU 3$ structure to include electric flux, and give a new example of an M5-brane wrapping a Kähler two-cycle and a Killing direction. We finish with a preliminary analysis of the generic $\SU 3$ and local $\SU 2$ structure cases.

\subsection{\texorpdfstring{\SU4}{SU(4)} holonomy and SLAG three-cycles\label{subsec:L=00003D0}}

Let us start by analysing the case where the Killing vector $L$ vanishes somewhere on $X$. As we will see, these solutions are always Minkowski. Using either the Fierz identities \eqref{eq:fierz} and \eqref{eq:Lsquared}, or the basis of spinors in Appendix \ref{par:spinors L=00003D0}, it is straightforward to check that for $L=0$ we have
\begin{equation}
S=0,\qquad K=0,\qquad P\lrcorner\bar{P}=2(1-\zeta^{2})=2\cos^{2}\alpha.\label{eq:S=00003D0 vectors}
\end{equation}
Note this means that all of the vectors vanish for $|\zeta|=1$, which give the cases with only chiral or antichiral spinors, and correspond to $\SU 4$ structures. As we will discuss, the Killing spinor equations then imply $m=0$ locally, but, since $m$ is constant, one must have $m=0$ on all of $X$. For $|\zeta|<1$, the one-form $P$ is non-zero and the differential condition for $S$ in \eqref{eq:S} implies that $m=0$. In conclusion, solutions with vanishing $L$ somewhere on $X$ are always Minkowski compactifications. Conversely, AdS$_{3}$ backgrounds have $\Vert L\Vert>0$ \emph{everywhere} -- $L$ is then a nowhere-vanishing Killing vector, as is expected for the R-symmetry vector field.\footnote{Alternatively, one can see this from \eqref{eq:Lsquared}, which implies that $L$ vanishes if and only if both $K$ and $S$ are zero. If $P$ is non-zero, then \eqref{eq:S} forces $m=0$. If $P$ also vanishes, all of the vectors are zero and the only solution is Minkowski times an $\SU 4$ structure manifold.}

From the expressions for $\beta$ and $g(S,\beta)$ in \eqref{eq:beta} and \eqref{eq:g}, it is simple to check this class of solutions satisfies
\begin{equation}
\beta=\sqrt{1-\zeta^{2}}\in[0,1),\qquad g(S,\beta)=|\zeta|.
\end{equation}
The results of Table \ref{tab:Stabiliser-group-of} imply that these solutions have $\SU 3$ structure, with an enhancement to $\SU 4$ for $|\zeta|=1$, corresponding to the rows $\op{int}\mathcal{R}^{\pm}$ and $\partial I^{\pm}$ respectively. We now give some examples of how known solutions fit within this class.

\subsubsection{Strict $\protect\SU 4$ structure\label{subsec:Strict--su4}}

An $\SU 4$ structure on an eight-manifold is equivalent to the existence of a real two-form $j_{\SU 4}$ and a complex four-form $\theta_{\SU 4}$, which are type $(1,1)$ and $(4,0)$ with respect to the almost complex structure defined by the pair. In an orthonormal frame $\{e^{a}\}$, these can be written as
\begin{equation}
j_{\SU 4}=e^{12}+e^{34}+e^{56}+e^{78},\qquad\theta_{\SU 4}=(e^{1}+\ii e^{2})\wedge\dots\wedge(e^{7}+\ii e^{8}),\label{eq:su4_tensors}
\end{equation}
from which we see they satisfy the algebraic conditions
\begin{equation}
\vol_{8}=\frac{1}{4!}j_{\SU 4}\wedge j_{\SU 4}\wedge j_{\SU 4}\wedge j_{\SU 4}=\frac{1}{2^{4}}\theta_{\SU 4}\wedge\bar{\theta}_{\SU 4}.
\end{equation}
An $\SU 4$ structure on $X$ corresponds to the case where the Killing spinors are both chiral or both antichiral. Focusing on the chiral case $\zeta=1$, in terms of the unit norm spinors in Appendix \ref{par:spinors L=00003D0}, it is simple to see that the general solution is
\begin{equation}
\chi_{1}=\epsilon_{1},\qquad\chi_{2}=\pm\epsilon_{3}.\label{eq:chi_su(4)}
\end{equation}
Note that swapping the sign of $\epsilon_{3}$ simply flips the complex structure, so that $(1,0)$-forms become $(0,1)$-forms, and so on. We pick the upper sign, $\chi_{2}=\epsilon_{3}$, in what follows.  Since both Killing spinors are chiral, all one-form spinor bilinears vanish identically and there are no invariant vectors, in agreement with \eqref{eq:S=00003D0 vectors}. As we now show, this reproduces the class of solutions in \cite{hep-th/9605053}, where the combination $\chi_{1}+\ii\chi_{2}$ is a complex chiral spinor (with its real and imaginary parts non-vanishing), corresponding to a (global) strict $\SU 4$ structure. The solutions are all Minkowski, and so there is no need for the existence of an R-symmetry Killing vector, consistent with the lack of vector fields in this case.

The various spinor bilinears of Table \ref{tab:bilinears} can be written in terms of the $\SU 4$-invariant tensors. Using the orthonormal frame in Appendix \ref{sec:An-orthonormal-frame}, it is simple to check that the non-trivial bilinears are
\begin{equation}
J=\omega=j_{\SU 4},\qquad\Phi=-\tfrac{1}{2}j_{\SU 4}\wedge j_{\SU 4},\qquad\Gamma=\theta_{\SU 4},
\end{equation}
and their Hodge duals.  Using these expressions, one finds that the differential conditions in Section \ref{subsec:Differential-and-algebraic} reduce to
\begin{equation}
f=3\dd\Delta,\qquad m=0,\qquad\dd(\ee^{3\Delta}j_{\SU 4})=0,\qquad\dd(\ee^{6\Delta}\theta_{\SU 4})=0,\qquad F=\star F.
\end{equation}
Note that $F$ is self-dual, so it must be type $(4,0)+(2,2)+(0,4)$ with respect to the almost complex structure. The remaining conditions from the algebraic Killing spinor equation in Appendix \ref{sec:Algebraic-Killing-spinor} are equivalent to
\begin{equation}
\theta_{\SU 4}\lrcorner F=0,\qquad j_{\SU 4}\lrcorner(F+\star F)=0.
\end{equation}
Together with the differential conditions, we have that $F$ is type $(2,2)$ and primitive. The full set of supersymmetry conditions is
\begin{equation}
\begin{gathered}f=3\dd\Delta,\qquad m=0,\qquad\dd(\ee^{3\Delta}j_{\SU 4})=0,\qquad\dd(\ee^{6\Delta}\theta_{\SU 4})=0,\\
F=F_{(2,2)},\qquad j_{\SU 4}\lrcorner F=0,
\end{gathered}
\label{eq:SU4_solution}
\end{equation}
with $F$ otherwise arbitrary. The metric on $X$ is conformal Calabi--Yau, as first derived in \cite{hep-th/9605053}. Since supersymmetry implies $m=0$ in this case, one cannot have a $(2,0)$ $\AdS 3$ background with spinors whose stabiliser enhances to $\SU 4$ somewhere on $X$. Only Minkowski compactifications can have spinors of this kind.

As we discussed in Section \ref{subsec:Bianchi-identities}, for $S=0$, supersymmetry does not automatically imply the equations of motion. Instead, we also have to impose the equation of motion for $f$ and the Bianchi identity for $F$, given in \eqref{eq:Bianchi (1,0)}. Provided the above supersymmetry conditions are satisfied, the equation of motion for $f$ reads
\begin{equation}
\begin{split}0=\ee^{-6\Delta}\dd(\ee^{6\Delta}\star f)+\tfrac{1}{2}F\wedge F & =3\dd\star\dd\Delta+18\dd\Delta\wedge\star\dd\Delta+\tfrac{1}{2}\Vert F\Vert^{2}\vol_{8}\\
 & =\left(-3\square_{8}\Delta-18\Vert\dd\Delta\Vert^{2}+\tfrac{1}{2}\Vert F\Vert^{2}\right)\vol_{8}.
\end{split}
\end{equation}
Comparing with the results of \cite{Martelli:2003ki}, we see that this is equivalent to the external part of the Einstein equation. As discussed in \cite{hep-th/9605053}, we also have to impose the Bianchi identity for $F$. Together, these extra conditions amount to a constraint on $F$ and the above differential equation for the warp factor $\Delta$.

\subsubsection{SLAG three-cycles in $\protect\SU 3$ structure manifolds\label{subsec:SLAG-three-cycles-in}}

Let us consider a more general class of Minkowski backgrounds where $\zeta$ is allowed to vary over $X$. Using $\zeta=\sin\alpha$ from \eqref{eq:S_zeta_def}, the non-zero bilinears are
\begin{equation}
\begin{gathered}P=\ii\cos\alpha(e^{7}+\ii e^{8}),\qquad\omega=j+\sin\alpha\,e^{78},\qquad J=\sin\alpha\,j+e^{78},\\
\varphi=\cos\alpha\re\theta,\qquad\Omega=\cos\alpha\,j\wedge(e^{7}+\ii e^{8}),\qquad\phi=\cos\alpha\im\theta,\\
\Phi=-\tfrac{1}{2}j\wedge j-\sin\alpha\,j\wedge e^{78},\qquad\Gamma=(\re\theta+\ii\sin\alpha\im\theta)\wedge(e^{7}+\ii e^{8}),
\end{gathered}
\end{equation}
where $(j,\theta,e^{7},e^{8})$ define a canonical $\SU 3$ structure in eight dimensions, as described in Appendix \ref{subsec:A-canonical-}. This corresponds to an $\SU 3$ structure that interpolates between $\SU 4$ structures of opposite chirality. If $\sin\alpha$ is constant over the internal manifold, this is known as an intermediate $\SU 3$ structure; if it varies, it is called a dynamic $\SU 3$ structure. One can show using the differential and algebraic conditions that $m=0$, while the electric flux is non-vanishing and given by
\begin{equation}
f=\ee^{-3\Delta}\dd(\ee^{3\Delta}\sin\alpha).
\end{equation}
The remaining conditions are
\begin{equation}
\begin{gathered}\dd(\ee^{3\Delta}P)=0,\qquad\dd(\ee^{3\Delta}\omega)=0,\\
\ee^{-3\Delta}\dd(\ee^{3\Delta}\varphi)=f\lrcorner\star\varphi+J\cdot F,\qquad\ee^{-6\Delta}\dd(\ee^{6\Delta}\phi)=\sin\alpha\,F-\star F,
\end{gathered}
\label{eq:gen_slag3}
\end{equation}
together with some algebraic constraints on $f$ and $F$ which can be derived from \eqref{eq:alg1}--\eqref{eq:P.f}. In particular, the one-form conditions imply the scalar conditions in this case, with the one-form conditions reducing to
\begin{equation}
\begin{gathered}F\wedge j\wedge(j+2\sin\alpha\,e^{78})=0,\qquad2\star\dd\alpha=\star F\wedge\im\theta,\\
\cos\alpha\,\star F\wedge j\wedge(e^{7}+\ii e^{8})=0,\qquad6\cos\alpha\star\dd\Delta=(\sin\alpha\star F+F)\wedge\im\theta.
\end{gathered}
\end{equation}
The differential condition for $P$ implies that both its real and imaginary parts are hypersurface orthogonal, so the metric on $X$ can be written as
\begin{equation}
\dd s^{2}(X)=\dd s^{2}(X_{6})+\frac{1}{\ee^{6\Delta}\cos^{2}\alpha}(\dd x^{2}+\dd y^{2}),
\end{equation}
where $X_{6}$ admits an $\SU 3$ structure, and we have introduced coordinates as $\ee^{3\Delta}P=\dd x+\ii\,\dd y$. Note that $P$ is not Killing, so $\Delta$ and $\alpha$ can depend on all coordinates on $X$.

For the special case where $\zeta\equiv\sin\alpha=0$, the conditions simplify to
\begin{equation}
\begin{gathered}f=0,\qquad m=0,\qquad\dd(\ee^{3\Delta}P)=0,\qquad\dd(\ee^{3\Delta}j)=0,\\
\ee^{-6\Delta}\dd(\ee^{6\Delta}\im\theta)=-\star F,\qquad\ee^{-3\Delta}\dd(\ee^{3\Delta}\re\theta)=(\tfrac{1}{2}\ii P\wedge\bar{P})\cdot F,
\end{gathered}
\end{equation}
with the simplified algebraic conditions equivalent to
\begin{equation}
\begin{gathered}F\wedge j\wedge j=0,\qquad\star F\wedge\im\theta=0,\\
\star F\wedge j\wedge(e^{7}+\ii e^{8})=0,\qquad6\star\dd\Delta=F\wedge\im\theta.
\end{gathered}
\end{equation}
These are the equations describing M5-branes wrapping SLAG three-cycles in eight-manifolds with an $\SU 3$ structure, with the differential condition for $\im\theta$ giving the calibration condition. These equations were first found in \cite[Section 5]{Martelli:2003ki} and re-derived in \cite{hep-th/0605146} from a wrapped probe-brane approach.\footnote{Solutions of this class were also given in \cite{hep-th/0012195}.} The conditions given in \eqref{eq:gen_slag3} should be thought of as a generalisation of these conditions to include M2-brane charge. Since $S$ vanishes for these solutions, one needs to check the Bianchi identity for $F$ and the equation of motion for $f$ in order to have a solution to the supergravity equations of motions. As shown in \cite{Martelli:2003ki} for the $\zeta=0$ case, this gives further differential constraints on the structure.

\paragraph{}

\subsection{\texorpdfstring{$\Gx2$}{G2} holonomy solutions\label{subsec:-holonomy-solutions}}

Consider the case where $\zeta=0$ and $|S|=1$, so that $S$ is a complex phase. These solutions have $\beta=1$ and so, from Table \ref{tab:Stabiliser-group-of}, admit a $\Gx 2$ structure. The relevant bilinears are
\begin{equation}
\begin{gathered}\zeta=0,\qquad S=\ee^{\ii\sigma},\qquad K=0,\qquad-\ii\bar{S}P=L=e^{8},\\
-\ii\bar{S}\Omega=\varphi=-\im\theta-j\wedge e^{7},\qquad\Phi=\bar{S}\star\Gamma=-L\lrcorner\star\varphi.
\end{gathered}
\end{equation}
It is straightforward to see that $L\lrcorner\varphi=L\lrcorner\Phi=0$ and
\begin{equation}
\tfrac{1}{7}\varphi\wedge\Phi\wedge L=\vol_{8}.
\end{equation}
From these constraints, we have a $\Gx 2$ structure in eight dimensions defined by $\varphi$ and $L$. Using the vanishing of various bilinears, the differential conditions in Section \ref{subsec:Differential-and-algebraic} reduce to
\begin{equation}
f=F=\dd\Delta=\dd S=m=0,\qquad\dd\varphi=\dd\Phi=\dd L=0.
\end{equation}
Thus the solutions are Minkowski with vanishing flux and constant warp factor. The phase $S$ is constant on $X$, and can be set to unity without loss of generality -- this corresponds to the case where the spinors $\chi_{i}$ have definite but opposite chirality globally on $X$. In terms of the unit norm spinors in Appendix \ref{subsec:Chiral-and-antichiral}, the Killing spinors can be taken to be
\begin{equation}
\chi_{1}=\epsilon_{1},\qquad\chi_{2}=\epsilon_{2}.
\end{equation}
The metric on $X$ is a product
\begin{equation}
\dd s^{2}(X)=\dd s^{2}(X_{7})+L\otimes L,
\end{equation}
where the metric on the seven-dimensional space $X_{7}$ transverse to $L$ has $\Gx 2$ holonomy. Since $L$ is a Killing vector, one can reduce along this direction to give a compactification of type IIA on a $\Gx 2$ holonomy manifold.

\subsection{AdS from wrapping Kähler four-cycles in \texorpdfstring{$\SU4$}{SU(4)} holonomy\label{subsec:AdS-from-wrapping}}

Let us consider a class of AdS solutions where $L$ is non-vanishing but $S=P=0$. These solutions have
\begin{equation}
|S|=\beta=0,
\end{equation}
and so fall within the case of $\op{int}I$ in Table \ref{tab:Stabiliser-group-of}. As we will see, this will give the conditions for supersymmetric AdS$_{3}$ solutions derived in \cite{hep-th/0608055}, which come from M5-branes wrapping Kähler four-cycles (holomorphic surfaces) in a Calabi--Yau fourfold.\footnote{Note that these solutions do not fall within the classes studied in \cite{hep-th/0605146}.} 

Using the spinors of Appendix \ref{subsec:S=00003D0} and again defining $\zeta=\sin\alpha$, the relevant bilinears are
\begin{equation}
\begin{gathered}K=\cos\alpha\,e^{8},\qquad L=\cos\alpha\,e^{7},\qquad P=0,\\
J=j+\sin\alpha\,e^{78},\qquad\omega=e^{78}+\sin\alpha\,j,\\
\varphi=\cos\alpha\,j\wedge e^{8},\qquad\Omega=\ii\cos\alpha\,\theta,\qquad\phi=-\cos\alpha\,j\wedge e^{7},\\
\Phi=-\tfrac{1}{2}j\wedge j-\sin\alpha\,j\wedge e^{78},\qquad\Gamma=\theta\wedge(e^{7}+\ii\sin\alpha\,e^{8}).
\end{gathered}
\end{equation}
We see these solutions admit an $\SU 3$ structure, in agreement with the results of Table \ref{tab:Stabiliser-group-of}. The differential and algebraic conditions then reduce to
\begin{equation}
\begin{gathered}\ee^{-3\Delta}\dd(\ee^{3\Delta}\sin\alpha)=f-4mK,\qquad\dd(\ee^{3\Delta}K)=0,\\
\ee^{-3\Delta}\dd(\ee^{3\Delta}L)=-2mJ-\tfrac{1}{2}\omega\lrcorner F+\tfrac{1}{2}J\lrcorner\star F,\qquad\ee^{-6\Delta}\dd(\ee^{6\Delta}\Omega)=-4m\Gamma,\\
\ee^{-6\Delta}\dd(\ee^{6\Delta}\phi)=-4m\Phi+\sin\alpha\,F-\star F,\qquad\ee^{-6\Delta}\dd(\ee^{6\Delta}\Phi)=-K\wedge F,\\
\dd\Delta=\tfrac{1}{3}\sin\alpha\,f-\tfrac{1}{6}F\lrcorner\star\phi,\qquad2m=\tfrac{1}{3}K\lrcorner f-\tfrac{1}{6}(\star\Phi)\lrcorner F,\\
\Omega\wedge\star F=\Omega\wedge F=0.
\end{gathered}
\label{eq:4in8}
\end{equation}
We have checked that these give the conditions for the backreaction of M5-branes wrapping Kähler four-cycles in a Calabi--Yau fourfold, as given in \cite[Appendix A]{hep-th/0608055}.\footnote{One subtlety in comparing the two sets of expressions is that the algebraic condition in \cite{hep-th/0608055} is given as $\im\theta\wedge F=0$. However, note that $L$ is a Killing vector that preserves the full solution including $F$. Then note that the differential conditions for $\Omega$ and $\sin\alpha$ above imply
\[
\mathcal{L}_{L}\theta=-4\ii m\theta.
\]
This means that $\theta$ is charge $-4m$ under the $\Uni 1$ generated by $L$. Since the full solution is invariant under the action of this $\Uni 1$, the supersymmetry conditions should hold after a $\Uni 1$ transformation, for which the algebraic condition would read
\[
(\cos4m\psi\,\im\theta-\sin4m\psi\,\re\theta)\wedge F=0,
\]
where $\psi$ is a coordinate along the fibre. This should hold for all $\psi$, which then implies that the stronger condition $F\wedge\Omega\equiv F\wedge\theta=0$ in \eqref{eq:4in8} must hold.} For example, the condition above which fixes the derivative of the warp factor in terms of $f$ and $F$ can be rewritten as
\begin{equation}
6\,\ee^{-\Delta}\dd(\ee^{\Delta})=2\sin\alpha\,f-\cos\alpha\,(j\wedge e^{7})\lrcorner\star F,
\end{equation}
which matches Equation (A.42) of \cite{hep-th/0608055} upon identifying $\ee^{\Delta}\equiv\omega$, $\sin\alpha\equiv-\cos2\beta$, $\cos\alpha\equiv-\sin2\beta$, $j\equiv J$, $e^{7}\equiv\hat{w}$, $f\equiv E_{1}$, and $F\equiv\ee^{-3\Delta}B_{4}$. Similarly, the differential equation for $\Omega$ can be expressed as
\begin{equation}
\ee^{-6\Delta}\dd(\ee^{6\Delta}\cos\alpha\im\theta)=4m(\re\theta\wedge e^{7}-\sin\alpha\im\theta\wedge e^{8}),
\end{equation}
which reproduces Equation (A.43) of \cite{hep-th/0608055} upon noting that $m_{\text{there}}=1/2$. Furthermore, upon setting the electric flux $f$ to zero, one recovers the supersymmetry conditions given in \cite{1508.04135}, corresponding to geometries which are sourced by M5-branes only. Since $S$ vanishes for this class of solutions, one needs to impose the Bianchi identity for $F$ and the equation of motion for $f$ in order to have a solution to the supergravity equations of motions~\cite{hep-th/0505230}.

\subsection{All purely electric solutions\label{subsec:All-purely-electric}}

As a special case of the previous subsection, one can consider solutions with non-zero electric flux, $f\neq0$, and the magnetic flux set to zero, $F=0$. In the AdS case, $m\neq0$, it is straightforward to show that the differential and algebraic conditions from the Killing spinor equations set $S=P=0$. One can then solve for the flux $f$, the one-form $K$ and the warp factor $\Delta$ in terms of the function $\zeta$ as\footnote{This identification relies only on an $N=1$ subsector of the Killing spinor equations, so it is not surprising that this matches \cite{Martelli:2003ki} after sending $\zeta\mapsto\sin\zeta$.}
\begin{equation}
f=\frac{3\dd\zeta}{1-\zeta^{2}},\qquad K=\frac{\dd\zeta}{2m},\qquad\ee^{-2\Delta}=1-\zeta^{2}.
\end{equation}
From Table \ref{tab:Stabiliser-group-of}, since $\beta=0$ and $|\zeta|\neq1$, these solutions admit an $\SU 3$ structure in the class $\op{int}I$. The remaining differential conditions are equivalent to
\begin{equation}
\ee^{-2\Delta}\dd(\ee^{2\Delta}L)=2mj,\qquad\dd(\ee^{2\Delta}j)=0,\qquad\ee^{-3\Delta}\dd(\ee^{3\Delta}\theta)=4\ii m\ee^{2\Delta}L\wedge\theta.
\end{equation}
Upon defining $\hat{\sigma}=\ee^{2\Delta}L$, $\hat{j}=\ee^{2\Delta}j$, $\hat{\theta}=\ee^{3\Delta}\theta$, the eleven-dimensional metric reads
\begin{equation}
\dd s_{11}^{2}=\frac{1}{1-\zeta^{2}}\left(\dd s^{2}(\AdS 3)+\frac{1}{4m^{2}}\frac{1}{1-\zeta^{2}}\dd\zeta\otimes\dd\zeta\right)+\dd s^{2}(X_{7}),
\end{equation}
where the metric defined by $(\hat{\sigma},\hat{j},\hat{\theta})$ on the seven-dimensional space $X_{7}$ is Sasaki--Einstein with $L$ as the Reeb vector field. The metric on $\dd s^{2}(\AdS 3)$ combines with the $\dd\zeta$ direction to give AdS$_{4}$ with radius $1/2m$; in particular, the change of variables $\sqrt{1-\zeta^{2}}=\cosh(2mr)$ gives AdS$_{4}$ foliated by copies of AdS$_{3}$.

In the Minkowski case, $m=0$, we again have $S=0$. Consistency of the algebraic conditions \eqref{eq:alg1} and \eqref{eq:L_cond} then requires $\zeta=\pm1$, and so we are back to the example of a strict $\SU 4$ structure as in Section \ref{subsec:Strict--su4}. The Killing spinors are both chiral or antichiral depending on the sign of $\zeta$, with the electric flux given by
\begin{equation}
f=\pm3\dd\Delta.
\end{equation}
After a conformal rescaling of the metric on $X$, one can take the resulting $\SU 4$ holonomy manifold to be a cone over a seven-dimensional Sasaki--Einstein manifold. Choosing the warp factor appropriately as a function of the radial coordinate on the cone, one again finds an $\AdS 4$ times Sasaki--Einstein solution, though now with the AdS$_{4}$ foliated by $\bR^{1,2}$~\cite{Martelli:2003ki}.

\subsection{AdS from wrapping co-associative cycles in \texorpdfstring{$\Gx2$}{G2} holonomy\label{subsec:AdS-from-wrapping-1}}

Let us consider the case where $L$ is non-vanishing but $\zeta=0$. As we saw in Section \ref{subsec:-holonomy-solutions}, restricting to $|S|=1$ forces us back to a Minkowski spacetime times an eight-manifold with $\Gx 2$ holonomy. Instead, we assume $|S|<1$ in what follows. From Table \ref{tab:Stabiliser-group-of}, there are then two cases to consider: $\beta=1$ or $\beta=|S|$. We consider $\beta=1$ in this subsection, corresponding to the $\op{int}D$ case with $\SU 3$ structure. This will give the conditions for supersymmetric AdS$_{3}$ solutions with vanishing electric flux derived in \cite{hep-th/0605146}, which come from the backreaction of M5-branes wrapping co-associative cycles in $\Gx 2$ holonomy manifolds.

Using the spinors of Appendix \ref{subsec:-and} and defining $S=\ee^{\ii\sigma}\sin\alpha$, it is sufficient to consider the bilinears
\begin{equation}
\begin{gathered}P=\sqrt{1-\sin^{2}\alpha\cos^{2}\sigma}e^{8}+\ii\frac{\cos\alpha\,e^{7}-\sin^{2}\alpha\cos\sigma\sin\sigma\,e^{8}}{\sqrt{1-\sin^{2}\alpha\cos^{2}\sigma}},\qquad Z+\ii L=\bar{S}\,P,\\
J=\frac{\ii}{2}P\wedge\bar{P},\qquad\omega=-\cos\alpha\,j,\\
\phi=\cos\alpha\re\tilde{\theta},\qquad\varphi=\im\tilde{\theta}+\frac{1}{\cos\alpha}j\wedge Z,\\
\Phi=-\tfrac{1}{2}j\wedge j+\frac{1}{\cos\alpha}\re\tilde{\theta}\wedge Z,
\end{gathered}
\label{eq:ads_coass_bilinears}
\end{equation}
where the one-form $K$ is zero and we have defined the real one-form $Z$ which lies in the $e^{78}$ plane and is orthogonal to $L$. We have also defined a rotated $\SU 3$ structure three-form $\tilde{\theta}$ as
\begin{equation}
\tilde{\theta}=\frac{(\sin\alpha\sin\sigma+\ii\cos\alpha)}{\sqrt{1-\sin^{2}\alpha\cos^{2}\sigma}}\theta,
\end{equation}
which corresponds to a phase rotation of the complex vielbein (and so this does not affect $L$, $Z$ or $j$). Note that for $S=0$, the one-form $L$ vanishes and the background is necessarily Minkowski. This reduces to the case of M5-branes wrapping SLAG three-cycles in $\SU 3$ structure manifolds discussed in Section \ref{subsec:SLAG-three-cycles-in}. For what follows, we assume $S\neq0$. 

Since $\zeta$ is zero by assumption and the one-form $K$ is vanishes, the differential condition on $\zeta$ implies that the electric flux vanishes, $f=0$. The remaining conditions reduce to
\begin{equation}
\begin{gathered}\ee^{-3\Delta}\dd(\ee^{3\Delta}S)=-4mP,\qquad\dd(\ee^{3\Delta}P)=0,\\
\ee^{-6\Delta}\dd(\ee^{6\Delta}J)=-L\lrcorner\star F,\qquad\ee^{-3\Delta}\dd(\ee^{3\Delta}\omega)=-L\lrcorner F,\\
\ee^{-3\Delta}\dd(\ee^{3\Delta}\varphi)=J\cdot F,\qquad\ee^{-6\Delta}\dd(\ee^{6\Delta}\Phi)=0.
\end{gathered}
\label{eq:coassociative}
\end{equation}
Using the derivative of $S$, one can show that $L$ and $Z$ can be written as
\begin{equation}
L=-\frac{1}{4m}\sin^{2}\alpha\,\dd\sigma,\qquad\ee^{-6\Delta}\dd(\ee^{6\Delta}\sin^{2}\alpha)=-8mZ.
\end{equation}
Using these expressions, or the derivatives of $S$ and $P$ directly, it is simple to show that $L\wedge\dd L=Z\wedge\dd Z=0$, and hence the vectors dual to $L$ and $Z$ are hypersurface orthogonal. Moreover, the vector dual to $L$ can be written as $L^{\sharp}=-4m\partial_{\sigma}$, so that $\sigma$ is a coordinate along the $\Uni 1$ fibre. Defining a new coordinate $\rho$ as
\begin{equation}
\rho=\ee^{3\Delta}\sin\alpha,
\end{equation}
the metric on $X$ can be expressed as
\begin{equation}
\begin{aligned}\dd s^{2}(X) & =\dd s^{2}(X_{6})+\frac{1}{\sin^{2}\alpha}\left(L\otimes L+\frac{1}{\cos^{2}\alpha}Z\otimes Z\right)\\
 & =\dd s^{2}(X_{6})+\frac{1}{16m^{2}}\ee^{-6\Delta}\left(\frac{1}{1-\ee^{-6\Delta}\rho^{2}}\dd\rho\otimes\dd\rho+\rho^{2}\dd\sigma\otimes\dd\sigma\right),
\end{aligned}
\end{equation}
where $X_{6}$ admits an $\SU 3$ structure defined by $(j,\tilde{\theta})$. Upon identifying $\ee^{-2\Delta}\equiv\lambda$ and noting that $m\equiv\tfrac{1}{2}m_{\text{there}}$, we recover the metric in \cite[Equation (6.1)]{hep-th/0605146}. Since $L$ is a Killing vector, the metric and warp factor are independent of the coordinate $\sigma$.

The magnetic flux is entirely determined by the differential conditions on $J$ and $\omega$ in \eqref{eq:coassociative}. To see this, first note that
\begin{equation}
-L\lrcorner\star F=\ee^{-6\Delta}\dd(\ee^{6\Delta}J)=\frac{\ii}{2}\ee^{-6\Delta}\dd(\ee^{6\Delta}P\wedge\bar{P})=0,
\end{equation}
where we used $\dd(\ee^{3\Delta}P)=0$ from \eqref{eq:coassociative}. This implies that $F$ is of the form $L\wedge\gamma$ for some three-form $\gamma$. Next, the differential condition for $\omega$ gives
\begin{equation}
\ee^{-3\Delta}\dd(\ee^{3\Delta}\cos\alpha\,j)=L\lrcorner F.
\end{equation}
Together with the result that the norm of $L$ is $\Vert L\Vert^{2}=\sin^{2}\alpha$, the magnetic flux is given by
\begin{equation}
F=\frac{1}{\sin^{2}\alpha}L\wedge\ee^{-3\Delta}\dd(\ee^{3\Delta}\cos\alpha\,j).\label{eq:F_coassociate}
\end{equation}
One can check that this matches the expression for the flux given in \cite[Equation (6.5)]{hep-th/0605146} after appropriately rescaling $j$ by the warp factor,\footnote{In our conventions, the eleven-dimensional metric is
\[
\dd s_{11}^{2}=\ee^{2\Delta}\left(\dd s^{2}(\AdS 3)+\dd s^{2}(X_{6})+e^{7}\otimes e^{7}+e^{8}\otimes e^{8}\right),
\]
and the $\SU 3$ structure forms $j$ and $\theta$ are given in an orthonormal frame for $\dd s^{2}(X_{6})$. In \cite{hep-th/0605146}, the $\SU 3$ structure is defined in terms of a frame for $\dd s^{2}(\mathcal{M}_{\SU 3})=\ee^{2\Delta}\dd s^{2}(X_{6})$, so that $j\equiv\ee^{-2\Delta}j_{\text{there}}$, and so on.} and noting that our one-form $L$ and their ``normalised'' $\text{S}^{1}$ volume are related as
\begin{equation}
\widehat{\vol}_{\text{S}^{1}}=-\frac{1}{\rho\lambda^{2}}L.
\end{equation}

The final two supersymmetry conditions in \cite{hep-th/0605146} come from the differential conditions for $\Phi$ and $\varphi$ in the final line of \eqref{eq:coassociative}. The differential condition on $\Phi$ reads
\begin{equation}
\dd\left(-\tfrac{1}{2}\ee^{6\Delta}j\wedge j+\ee^{6\Delta}\frac{1}{\cos\alpha}\re\tilde{\theta}\wedge Z\right)=0.\label{eq:Phi}
\end{equation}
The differential condition on $\varphi$ is
\begin{equation}
\ee^{-3\Delta}\dd\left(\ee^{3\Delta}\im\tilde{\theta}+\ee^{3\Delta}\frac{1}{\cos\alpha}j\wedge Z\right)=\frac{1}{\sin^{2}\alpha}Z\wedge\imath_{L}F.
\end{equation}
where we used $(L\wedge Z)\cdot F=-Z\wedge\imath_{L}F$ since $L\wedge F=0$. Inserting the expression for $F$ in \eqref{eq:F_coassociate} and rearranging, one finds this is equivalent to
\begin{equation}
\dd\left(\ee^{3\Delta}\im\tilde{\theta}+\frac{1}{\sin^{2}\alpha\cos\alpha}Z\wedge\ee^{3\Delta}j\right)=0.\label{eq:varphi}
\end{equation}
To make contact with the remaining conditions in \cite{hep-th/0605146}, we note that their one-form $\hat{\rho}$ is defined as
\begin{equation}
\hat{\rho}=\frac{\ee^{-2\Delta}\dd\rho}{4m\sqrt{1-\ee^{-6\Delta}\rho^{2}}},\qquad Z\equiv-\ee^{-\Delta}\sin\alpha\cos\alpha\,\hat{\rho}.
\end{equation}
With the identification of $\lambda$, $\phi$ and $m$ given above, and the rescaling of the $\SU 3$ structure forms by the warp factor, one finds that \eqref{eq:Phi} and \eqref{eq:varphi} exactly reproduce Equations (6.4) and (6.3) of \cite{hep-th/0605146} respectively.

Since the electric flux $f$ vanishes for this class of backgrounds, these geometries are sourced by M5-branes alone. As we discussed in Section \ref{subsec:Bianchi-identities}, for $S\neq0$, the Bianchi identity for $F$ should be implied by the supersymmetry conditions. Indeed, the explicit expression for $F$ in \eqref{eq:F_coassociate} and the conditions
\begin{equation}
\dd(\ee^{6\Delta}L)=\frac{8m}{\sin^{2}\alpha}\ee^{6\Delta}L\wedge Z,\qquad\dd(\ee^{6\Delta}\sin^{2}\alpha)=-8m\ee^{6\Delta}Z,
\end{equation}
which can be derived from the $S$ and $P$ derivatives, automatically imply $\dd(\ee^{3\Delta}F)=0$.

\subsection{Kähler two-cycles in \texorpdfstring{$\SU3$}{SU(3)} structure manifolds\label{subsec:New-example}}

Let us again consider the case where $L$ is non-vanishing but $\zeta=0$. These differ from the previous subsection as they will have
\begin{equation}
\beta=|S|,
\end{equation}
and not $\beta=1$. For $|S|=0$, these solutions live on the locus $\op{int}I$ at $\zeta=0$, and are captured by the solution in Section \ref{subsec:AdS-from-wrapping} in the $\zeta\to0$ limit. Thanks to this, we assume $|S|\neq0$ in what follows -- the solutions then correspond to the locus $\op{int}D\backslash\{0\}$ in Table \ref{tab:Stabiliser-group-of}. As we will show, these solutions are always Minkowski with vanishing electric flux, and correspond to M5-branes wrapping a Kähler two-cycle and a circle defined by the Killing direction $L$. To the author's knowledge, these solutions are new -- in particular, they generalise those in \cite{hep-th/0605146} where Kähler two-cycles were considered, which led to $\mathbb{R}^{1,3}$ solutions.

We do not immediately give all the bilinears in terms of the $\SU 3$ invariant forms, since there will be a simplification after considering the supersymmetry conditions. However, we note that
\begin{equation}
P=\ii SL,\qquad J=\sqrt{1-|S|^{2}}j,\qquad\omega=-K\wedge L,\qquad\phi=-L\wedge J,
\end{equation}
where $K$ and $L$ lie the in the $e^{78}$ plane, and their norms are $\Vert L\Vert^{2}=1$ and $\Vert K\Vert^{2}=1-|S|^{2}$. The relevant differential conditions are
\begin{equation}
\begin{gathered}4mK=f,\qquad\ee^{-3\Delta}\dd(\ee^{3\Delta}S)=-4\ii mSL.\\
\dd(\ee^{3\Delta}K)=0,\qquad\dd(\ee^{3\Delta}P)=0,\\
\ee^{-6\Delta}\dd(\ee^{6\Delta}J)=f\wedge\omega-L\lrcorner\star F,\qquad\ee^{-3\Delta}\dd(\ee^{3\Delta}\omega)=-L\lrcorner F,\\
\ee^{-6\Delta}\dd(\ee^{6\Delta}\phi)=-4m\Phi-\star F,\qquad\ee^{-6\Delta}\dd(\ee^{6\Delta}\Omega)=-4m\Gamma+S\,F,\\
\ee^{-6\Delta}\dd(\ee^{6\Delta}\Phi)=-K\wedge F.
\end{gathered}
\label{eq:not_coassociative}
\end{equation}
Since $\zeta=0$, the differential condition for $\zeta$ fixes
\begin{equation}
f=4mK.
\end{equation}
The differential conditions for $S$ and $P$ then give
\begin{equation}
0=\dd(\ee^{3\Delta}SL)=\ee^{3\Delta}S\,\dd L,
\end{equation}
so that either $S=0$ or $\dd L=0$. Since we assume $S\neq0$, $L$ must be closed. Furthermore, the differential condition on $K$ implies $K\wedge\dd K=0$, and so both $L$ and $K$ are hypersurface orthogonal.\footnote{The differential condition on $L$ contains the same information as the one-form identity in \eqref{eq:alg1}:
\[
\dd\Delta+\tfrac{1}{6}(L\wedge J)\lrcorner\star F=0,
\]
which fixes the derivative of the warp factor. This is implied by the conditions we give in the main text.} Given that $\omega=-K\wedge L$ and $f$ lies along the $K$ direction, $f\wedge\omega$ vanishes. The differential conditions for $J$ and $\omega$ then simplify to
\begin{equation}
\ee^{-6\Delta}\dd(\ee^{6\Delta}J)=-L\lrcorner\star F,\qquad\ee^{-3\Delta}\dd(\ee^{3\Delta}\omega)=-L\lrcorner F,
\end{equation}
which completely fix $F$. Given the form of $\omega$ in terms of $K$ and $L$, one also finds that
\begin{equation}
\dd(\ee^{3\Delta}\omega)=0,
\end{equation}
and so $F$ has no component along $L$, $L\lrcorner F=0$. The magnetic flux can then be written as
\begin{equation}
\star F=-L\wedge\ee^{-6\Delta}\dd(\ee^{6\Delta}\sqrt{1-|S|^{2}}j)\equiv\ee^{-6\Delta}\dd(\ee^{6\Delta}\sqrt{1-|S|^{2}}L\wedge j).\label{eq:new_calibration}
\end{equation}

Note that $\star F$ also appears in the condition for $\phi$:
\begin{equation}
\begin{aligned}\ee^{-6\Delta}\dd(\ee^{6\Delta}\phi) & =(L\lrcorner\star F)\wedge L=-\star F\\
 & \equiv-4m\Phi-\star F,
\end{aligned}
\end{equation}
where we used the derivatives of $J$ and $L$. Since $\Phi$ does not vanish identically, for consistency we must have $m=0$, so that the solution is Minkowski with vanishing electric flux. We can then interpret \eqref{eq:new_calibration} as giving the calibration condition for an M5-brane wrapping a holomorphic curve (a Kähler two-cycle) together with the $L$ direction in $X$. With $m=0$, the differential condition for $S$ implies that $\ee^{3\Delta}S$ is constant on $X$:
\begin{equation}
\dd(\ee^{3\Delta}S)=0.
\end{equation}
This means one has
\begin{equation}
\dd(\ee^{3\Delta}|S|)=0,\qquad\dd\sigma=0,
\end{equation}
where $S=|S|\ee^{\ii\sigma}$. Since a constant phase rotation of the complex spinor $\chi$ will still solve the Killing spinor equations, the constant phase in $S$ can always be set to $\sigma=0$, so that $S=|S|$.\footnote{Using the observation that constant shifts of the warp factor can always be absorbed by a constant rescaling of the Minkowski metric on $\mathbb{R}^{1,2}$, we are also free to set $\ee^{3\Delta}|S|=1$.} With this choice, the spinor bilinears are given by
\begin{equation}
\begin{gathered}K=\sqrt{1-S^{2}}e^{8},\qquad L=e^{7},\qquad P=\ii SL,\\
J=\sqrt{1-S^{2}}j,\qquad\omega=L\wedge K,\\
\varphi=S\re\theta+j\wedge e^{8},\qquad\phi=-L\wedge J,\qquad\Omega=\ii(\re\theta+\ii\sqrt{1-S^{2}}\im\theta)+\ii Sj\wedge e^{8},\\
\Phi=-\tfrac{1}{2}j\wedge j-S\im\theta\wedge e^{8},\qquad\Gamma=(\re\theta+\ii\sqrt{1-S^{2}}\im\theta)\wedge e^{7}-Sj\wedge e^{78}.
\end{gathered}
\label{eq:kahler_2}
\end{equation}
From the differential conditions on the three- and four-forms, it is straightforward to show the remaining conditions are
\begin{equation}
\begin{aligned}\ee^{-6\Delta}\dd(\ee^{6\Delta}\sqrt{1-S^{2}}\im\theta) & =-S\,F,\\
\ee^{-6\Delta}\dd(\ee^{6\Delta}\re\theta+\ee^{6\Delta}Sj\wedge e^{8}) & =0,\\
\ee^{-6\Delta}\dd\left(-\ee^{6\Delta}\tfrac{1}{2}j\wedge j-\ee^{6\Delta}S\im\theta\wedge e^{8}\right) & =-\sqrt{1-S^{2}}e^{8}\wedge F.
\end{aligned}
\end{equation}
Notice that the first of these implies the Bianchi identity for $F$, as we expect for $(2,0)$ solutions with $S\neq0$.

To write the metric, we introduce coordinates as
\begin{equation}
L=\dd\psi,\qquad\ee^{3\Delta}K=\dd y.
\end{equation}
The eleven-dimensional metric is then
\begin{equation}
\dd s^{2}=\ee^{2\Delta}\left(\dd s^{2}(\mathbb{R}^{1,2})+\dd s^{2}(X_{6})+\dd\psi^{2}+\frac{1}{1-S^{2}}\ee^{-6\Delta}\dd y\otimes\dd y\right),
\end{equation}
where the six-manifold $X_{6}$ admits an $\SU 3$ structure defined by $(j,\theta)$, and the metric is independent of the coordinate $\psi$ since $L$ is Killing. Given the calibration condition in \eqref{eq:new_calibration}, we interpret this as the geometry sourced by an M5-brane wrapping a Kähler two-cycle in $X_{6}$ together with the circle parametrised by $\psi$. 

If we send the radius of the circle to infinity, the solution becomes four-dimensional Minkowski space times a seven-manifold:
\begin{equation}
\dd s^{2}=\ee^{2\Delta}\left(\dd s^{2}(\mathbb{R}^{1,3})+\dd s^{2}(X_{6})+\frac{1}{1-S^{2}}\ee^{-6\Delta}\dd y\otimes\dd y\right).
\end{equation}
In the $S\to0$ limit, our conditions match those given in \cite{hep-th/0605146}, which describe an M5-brane wrapping a Kähler two-cycle in an $\SU 3$ holonomy manifold. In particular, our conditions simplify to
\begin{equation}
\tfrac{1}{2}\ee^{-6\Delta}\dd(\ee^{6\Delta}j\wedge j)=K\wedge F,\qquad\ee^{-6\Delta}\dd(\ee^{6\Delta}\theta)=0,\qquad\ee^{-6\Delta}\dd(\ee^{6\Delta}j)=-\imath_{L}\star F,
\end{equation}
which reproduce Equations (3.30)--(3.33) of \cite{hep-th/0605146} upon identifying $K\equiv-v$, $\star_{7}\equiv\imath_{L}\star$, and noting the different warp factors in the metric ansatz.

\subsection{\texorpdfstring{$\SU3$}{SU(3)} structure with \texorpdfstring{$\zeta\neq0$}{zeta =/= 0}, \texorpdfstring{$S\neq0$}{S =/= 0}\label{subsec:Generic-case-with}}

Consider the case of an $\SU 3$ structure with $\zeta$ and $S$ non-zero. We do not give a full analysis of this system, but are instead content to point out some salient features. The one-forms $K$, $L$ and $P$ are non-zero for generic values of $\zeta$ and $S$, however the kernel distribution of the four one-forms $(K,L,\re P,\im P)$ should be rank six for the transverse space to admit an $\SU 3$ structure. Thus, there must be two relations between them (or one complex relation). Since the Fierz identities \eqref{eq:fierz} imply that the one-forms $K$ and $L$ are orthogonal, $K\lrcorner L=0$, we can choose $P$ to be a sum of $K$ and $L$. The Fierz identities then imply that $P$ can be written as
\begin{equation}
P=S\left(-\zeta\frac{K}{\Vert K\Vert^{2}}+\ii\frac{L}{\Vert L\Vert^{2}}\right).\label{eq:P_SU3}
\end{equation}
This fixes the norm of $\re P$ in terms of the other quantities, and hence restricts the allowed values of the scalar $\beta$. Using the Fierz identities again, one finds that $\zeta$, $|S|$ and $\beta$ are related by
\begin{equation}
\zeta=\pm g(S,\beta),\label{eq:zeta-2}
\end{equation}
with $\beta$ constrained to the interval $(|S|,1)$ (so that $g(S,\beta)$ and hence $\zeta$ is real). Thus, we are on the locus $\op{int}\mathcal{R}^{\pm}$ of Table \ref{tab:Stabiliser-group-of}.\footnote{Recall that our other example of a solution in this class is the Minkowski background from M5-branes wrapped on SLAG three-cycles in Section \ref{subsec:SLAG-three-cycles-in}.} It is also straightforward to show that \eqref{eq:zeta-2} can be inverted to give
\begin{equation}
\beta=\sqrt{f_{\mp}(\zeta,S)},\label{eq:beta_su3}
\end{equation}
where $f_{\pm}$ are the roots of \eqref{eq:fpm}, with the norms of $K$ and $L$ fixed to
\begin{equation}
\Vert L\Vert^{2}=f_{\pm}(\zeta,S),\qquad\Vert K\Vert^{2}=f_{\pm}(\zeta,S)-|S|^{2}.
\end{equation}
From the discussion in Section \ref{sec:Analysis-of--structures}, since we are on the locus $\op{int}\mathcal{R}^{\pm}$, we have $0<|S|^{2}<f_{-}<f_{+}<1$, and hence $L$ and $K$ are non-zero. For what follows, we take the upper sign.

The scalar condition \eqref{eq:zeta} determines the electric flux as
\begin{equation}
f=\ee^{-3\Delta}\dd(\ee^{3\Delta}\zeta)+4mK,
\end{equation}
while the conditions on $S$ and $P$ from \eqref{eq:S} and \eqref{eq:P} imply
\begin{equation}
\dd\bigl(\zeta\Vert K\Vert^{-2}K\bigr)=0,\qquad\dd\bigl(\Vert L\Vert^{-2}L\bigr)=0.
\end{equation}
With this in mind, let us assume that $m\neq0$. Since $P$ can be written in terms of $K$ and $L$ as in \eqref{eq:P_SU3}, the differential condition for $S$ in \eqref{eq:S} implies we can integrate the above equations to find
\begin{equation}
K=\frac{1}{4m\zeta}\Vert K\Vert^{2}\dd\rho,\qquad L=-\frac{1}{4m}\Vert L\Vert^{2}\dd\sigma,
\end{equation}
where we have expressed $S=|S|\ee^{\ii\sigma}$ in terms of its magnitude and phase, and defined the coordinate $\rho=\log(\ee^{3\Delta}|S|)$. Since $\dd(S^{-1}P)=0$, both $K$ and $L$ are hypersurface orthogonal and define an integrable product structure, so the full metric is diagonal in $\dd\rho$ and $\dd\sigma$. The vectors dual to $K$ and $L$ are
\begin{equation}
K^{\sharp}=4m\zeta\,\partial_{\rho},\qquad L^{\sharp}=-4m\partial_{\sigma},
\end{equation}
and the one-form $P$ is simply
\begin{equation}
P=-\frac{S}{4m}(\dd\rho+\ii\,\dd\sigma).\label{eq:P_su3-1}
\end{equation}
The unwarped metric on $X$ can then be written as
\begin{equation}
\dd s^{2}(X)=\dd s^{2}(X_{6})+\frac{\Vert L\Vert^{2}}{16m^{2}}\dd\sigma^{2}+\frac{\Vert K\Vert^{2}}{16m^{2}\zeta^{2}}\dd\rho^{2},
\end{equation}
where $X_{6}$ admits an $\SU 3$ structure and, since $L$ is a Killing vector, the metric is independent of $\sigma$. Furthermore, since $K$ is uncharged under $L$, $\Vert K\Vert^{2}$ and hence $\zeta$ are independent of $\sigma$ too. This product structure also allows us to decompose the exterior derivative as
\begin{equation}
\dd=\dd_{6}+\dd\rho\wedge\partial_{\rho}+\dd\sigma\wedge\partial_{\sigma}.
\end{equation}
Using this, the differential condition on $K$ in \eqref{eq:K} reduces to
\begin{equation}
\dd_{6}\bigl(\ee^{3\Delta}\zeta^{-1}\Vert K\Vert^{2}\bigr)=0,\label{eq:K_cond}
\end{equation}
so that $\ee^{3\Delta}\zeta^{-1}\Vert K\Vert^{2}$ is a function of $\rho$ alone. Moreover, this relates the $\dd_{6}$ derivatives of the warp factor and the scalar $\zeta$ as
\begin{equation}
\dd_{6}\Delta=\frac{1-|S|^{2}+\zeta^{2}}{3(1+|S|^{2}-\zeta^{2})}\dd_{6}\log\zeta.
\end{equation}

For notational convenience, let us define
\begin{equation}
\zeta_{\pm}=\sqrt{(1+\zeta)^{2}-|S|^{2}}\pm\sqrt{(1-\zeta)^{2}-|S|^{2}},
\end{equation}
which satisfy $\zeta_{+}\zeta_{-}=4\zeta$. The norms of $L$ and $K$ can then be expressed as
\begin{equation}
\Vert L\Vert^{2}=1-\tfrac{1}{4}\zeta_{-}^{2},\qquad\Vert K\Vert^{2}=\tfrac{1}{4}\zeta_{+}^{2}(1-\tfrac{1}{4}\zeta_{-}^{2}),
\end{equation}
while the two-form bilinears are
\begin{equation}
\omega=\tfrac{1}{2}(\zeta_{-}j+\zeta_{+}e^{78}),\qquad J=\tfrac{1}{2}(\zeta_{+}j+\zeta_{-}e^{78}),
\end{equation}
where $e^{78}$ is fixed to
\begin{equation}
K\wedge L=\tfrac{1}{2}(\zeta\zeta_{-}-\zeta_{+})e^{78}.
\end{equation}
The two-form conditions from Section \ref{subsec:Differential-and-algebraic} then fix the magnetic flux in terms of the derivatives of $\omega$ and $J$ as
\begin{equation}
\begin{aligned}\imath_{L}F & =-\ee^{-3\Delta}\dd(\ee^{3\Delta}\omega),\\
\imath_{L}\star F & =f\wedge\omega-\ee^{-6\Delta}\dd(\ee^{6\Delta}J).
\end{aligned}
\label{eq:su_3_F-2-1}
\end{equation}
We can use the product structure to decompose the magnetic flux as
\begin{equation}
F=F_{4}+F_{3}^{K}\wedge\hat{K}+F_{3}^{L}\wedge\hat{L}+F_{2}^{KL}\wedge\hat{K}\wedge\hat{L},
\end{equation}
where $\hat{K}=K/\Vert K\Vert$ and $\hat{L}=L/\Vert L\Vert$ are unit one-forms and $F_{4}$, $F_{3}^{K}$, $F_{3}^{L}$ and $F_{2}^{KL}$ are four-, three- and two-forms along $X_{6}$. Defining a Hodge star on $X_{6}$ as $\star_{6}=(\hat{L}\wedge\hat{K})\lrcorner\star$,\footnote{One has the following identities for the Hodge star
\[
\begin{aligned}\star F_{4} & =((\hat{K}\wedge\hat{L})\lrcorner\star F_{4})\wedge\hat{K}\wedge\hat{L}, & \star(F_{3}^{L}\wedge\hat{L}) & =((\hat{L}\wedge\hat{K})\lrcorner\star F_{3}^{L})\wedge\hat{K},\\
\star(F_{3}^{K}\wedge\hat{K}) & =((\hat{K}\wedge\hat{L})\lrcorner\star F_{3}^{K})\wedge\hat{L}, & \star(F_{2}^{KL}\wedge\hat{K}\wedge\hat{L}) & =((\hat{L}\wedge\hat{K})\lrcorner\star F_{2}^{KL}).
\end{aligned}
\]
} the dual of the flux can be written as
\begin{equation}
\star F=\star_{6}F_{4}\wedge\hat{K}\wedge\hat{L}+\star_{6}F_{3}^{K}\wedge\hat{L}+\star_{6}F_{3}^{L}\wedge\hat{K}+\star_{6}F_{2}^{KL}.
\end{equation}
One can then evaluate \eqref{eq:su_3_F-2-1} to find
\begin{equation}
\begin{aligned}2\Vert L\Vert F_{3}^{L} & =\ee^{-3\Delta}\dd_{6}(\ee^{3\Delta}\zeta_{-}j),\\
\frac{1}{2m\zeta}\Vert L\Vert\Vert K\Vert F_{2}^{KL} & =\ee^{-3\Delta}\partial_{\rho}(\ee^{3\Delta}\zeta_{-}j),\\
-2\Vert L\Vert\star_{6}F_{3}^{K} & =\ee^{-3\Delta}\zeta_{-}\dd_{6}(\ee^{3\Delta}\zeta)\wedge j-\ee^{-6\Delta}\dd_{6}(\ee^{6\Delta}\zeta_{+}j),\\
-\frac{\Vert L\Vert\Vert K\Vert}{2m\zeta}\star_{6}F_{4} & =\left(2\Vert L\Vert\Vert K\Vert(1-3\partial_{\rho}\Delta)+\zeta_{-}\partial_{\rho}\zeta-\partial_{\rho}\zeta_{+}\right)j-\zeta_{+}\ee^{-3\Delta}\partial_{\rho}(\ee^{3\Delta}j).
\end{aligned}
\end{equation}
We see that all of the components of the magnetic flux are fixed by (derivatives of) $\zeta$, $S$ and $j$. 

Consider taking $\imath_{L}$ of the three-form differential conditions \eqref{eq:varphi-1}--\eqref{eq:Omega}. Using that the contractions are given by
\begin{equation}
\imath_{L}\varphi=0,\qquad\imath_{L}\phi=\tfrac{1}{2}(\zeta\zeta_{-}-\zeta_{+})j,\qquad\imath_{L}\Omega=\tfrac{1}{2}\zeta_{-}S\,j,
\end{equation}
and that $\varphi$ and $\phi$ are uncharged under $L$, while $\Omega$ is charge $-4m$, it is straightforward to check that the $\phi$ and $\Omega$ conditions are automatically satisfied given the zero-, one- and two-form conditions, while the derivative of $\varphi$ gives a constraint on the flux:
\begin{equation}
0=\imath_{L}(f\lrcorner\star\varphi)+\imath_{L}(J\cdot F).
\end{equation}
One can continue in this fashion and analyse the remaining differential conditions -- these will relate the torsion of the $\SU 3$ structure (in the form of the derivatives of $\re\theta$ and $\im\theta$) to the flux, and impose the Bianchi identity (since $S\neq0$). These conditions will be complicated and we do not expect them to have a clean geometric interpretation, other than as an $\SU 3$ structure in eight dimensions with all torsion classes turned on. We leave a full analysis of this system to future work.

\subsection{\texorpdfstring{$\SU2$}{SU(2)} structure with \texorpdfstring{$\zeta=0$}{zeta = 0}\label{sec:Examples-with-su2}}

As our final example, from Figure \ref{fig:Illustrations-of-the}, we see that there are backgrounds with a local $\SU 2$ structure which have $\zeta=0$, with $S$ and $\beta$ as free parameters. To the author's knowledge, backgrounds of this kind have not appeared in the literature. Using the orthonormal frame in Appendix \ref{sec:An-orthonormal-frame-1}, it is simple to check that the four one-forms are linearly independent away from $|S|=1$, with norms
\begin{equation}
\begin{aligned}\Vert L\Vert^{2} & =\tfrac{1}{2}(1+|S|^{2}), & \Vert K\Vert^{2} & =\tfrac{1}{2}(1-|S|^{2}),\\
\Vert\re P\Vert^{2} & =\tfrac{1}{2}(1-|S|^{2}\cos2\sigma), & \Vert\im P\Vert^{2} & =\tfrac{1}{2}(1+|S|^{2}\cos2\sigma),
\end{aligned}
\end{equation}
where $S=|S|\ee^{\ii\sigma}$. We see $L$ is nowhere-vanishing on the $\SU 2$ locus and so defines a transverse foliation of $X$. Using the Fierz identities in \eqref{eq:fierz}, one can check that $L$ is not orthogonal to $P$. We then define two new one-forms, $Y$ and $Z$, which lie in the plane defined by $(\re P,\im P)$:
\begin{equation}
\begin{aligned}Y & =|S|\cos\sigma\re P+|S|\sin\sigma\im P,\\
Z & =|S|\sin\sigma\re P-|S|\cos\sigma\im P+\frac{|S|^{2}}{\Vert L\Vert^{2}}L.
\end{aligned}
\end{equation}
The one-forms $(K,L,Y,Z)$ are mutually orthogonal and non-zero away from $|S|=1$, and so define a rank-four distribution on $X$. One can then define an orthonormal frame via $E^{1}=K/\Vert K\Vert$, and so on, so that the unwarped metric on the $\SU 2$ locus on the eight-manifold $X$ may be written as
\begin{equation}
\dd s^{2}(X)=\dd s^{2}(X_{4})+\sum_{i=1}^{4}E^{i}\otimes E^{i},
\end{equation}
where $X_{4}$ admits an $\SU 2$ structure. We can then analyse the supersymmetry conditions in turn, starting with the differential conditions on the scalar bilinears.

The scalar condition on $\zeta$ in \eqref{eq:zeta} fixes the electric flux to
\begin{equation}
f=4mK.\label{eq:f_su2}
\end{equation}
The differential condition on $S$ in \eqref{eq:S} can then be used to solve for the one-form $Y$ as
\begin{equation}
Y=-\frac{1}{4m}\ee^{-6\Delta}\rho\,\dd\rho,\label{eq:Y}
\end{equation}
where we have defined $\rho=\ee^{3\Delta}|S|$, which will serve as a coordinate on $X$. Similarly, one can show that a certain combination of $Z$ and $L$ is exact:
\begin{equation}
\frac{Z}{|S|^{2}}-\frac{L}{\Vert L\Vert^{2}}=\frac{1}{4m}\dd\sigma.
\end{equation}

Since $L$ is Killing, we can define an adapted coordinate $\psi$ via
\begin{equation}
L^{\sharp}=-4m\partial_{\psi},
\end{equation}
where the metric has no explicit $\psi$ dependence. Note that from \eqref{eq:S_deriv} the scalar bilinear $S$ should be charge $-4m$ under $L$. We can solve this condition by defining a new coordinate $\tau$ such that
\begin{equation}
S=|S|\ee^{\ii(\psi+\tau)},
\end{equation}
where $\imath_{L}\dd\tau=0$. The one-form $L$ is then given in coordinates by
\begin{equation}
L=-\frac{1}{4m}\Vert L\Vert^{2}(\dd\psi+\eta),
\end{equation}
where $\eta$ is a connection one-form which is basic with respect to $L$, i.e.~$\mathcal{L}_{L}\eta=\imath_{L}\eta=0$. Using this parametrisation, the one-form $Z$ is given by
\begin{equation}
Z=\frac{1}{4m}|S|^{2}(\dd\tau-\eta).
\end{equation}

Moving on to the one-forms, the differential condition on $P$ in \eqref{eq:P} is automatically satisfied given the above parametrisation in terms of the coordinates $(\psi,\rho,\tau)$. The condition on $K$ in \eqref{eq:K} implies that it is hypersurface orthogonal, and so defines an integrable almost product structure. As such, we can introduce a coordinate $w$ for which
\begin{equation}
K=\ee^{-3\Delta}\dd w.
\end{equation}
Similarly, from \eqref{eq:Y} we have $Y\wedge\dd Y=0$, and so one can choose coordinates where $\eta$ has no $\dd\psi$, $\dd w$ or $\dd\rho$ components, and the metric is diagonal in $\dd w$ and $\dd\rho$.\footnote{From Section \ref{sec:Analysis-of--structures}, we expect there to be a rank-three product structure on $X$. This is spanned by $K$, $Y$ and $\tilde{Z}=Z-\frac{|S|^{2}}{\Vert L\Vert^{2}}L$ -- these one-forms are mutually orthogonal and satisfy $K\wedge\dd K=Y\wedge\dd Y=\tilde{Z}\wedge\dd\tilde{Z}$. In practice, it proves simpler to work with $Z$ rather than $\tilde{Z}$, since it is also orthogonal to $L$.}

Having fixed the one-forms, we have solved the $S$, $P$, $K$, and Killing vector conditions, with the electric flux fixed by \eqref{eq:f_su2}. We then have to impose the $L$ condition from \eqref{eq:L}. Using the algebraic conditions in Appendix \ref{sec:Algebraic-Killing-spinor}, the $L$ condition can be written as
\begin{equation}
\dd L=4mJ-f\lrcorner\varphi-\omega\lrcorner F,
\end{equation}
where
\begin{align}
4mJ-f\lrcorner\varphi & =\frac{1}{4m}\ee^{-6\Delta}\rho\,\dd\rho\wedge(\dd\tau+\dd\psi).
\end{align}
Given a magnetic flux $F$, this fixes the curvature of the connection one-form $\eta$ to
\begin{align}
\Vert L\Vert^{2}\dd\eta & =\frac{16m^{2}}{|S|^{2}}Y\wedge Z+\frac{4m}{\Vert L\Vert^{2}}\imath_{L}\bigl(L\wedge(\omega\lrcorner F)\bigr).\label{eq:connection}
\end{align}

Recall that the two-form conditions are
\begin{equation}
\begin{aligned}\ee^{-6\Delta}\dd(\ee^{6\Delta}J) & =f\wedge\omega-L\lrcorner\star F,\\
\ee^{-3\Delta}\dd(\ee^{3\Delta}\omega) & =-L\lrcorner F,
\end{aligned}
\label{eq:2-form_su2}
\end{equation}
with the relevant bilinears given in terms of the canonical $\SU 2$ structure of Appendix \ref{sec:An-orthonormal-frame-1} by
\begin{align}
\omega & =\tfrac{1}{2}\sqrt{1-|S|^{2}}(-j_{\SU 2}-\im\theta_{\SU 2})-K\wedge\left(\frac{L}{\Vert L\Vert^{2}}+\frac{2}{1-|S|^{2}}Z\right),\\
J & =\tfrac{1}{2}\sqrt{1-|S|^{2}}(-j_{\SU 2}+\im\theta_{\SU 2})+Y\wedge\left(\frac{L}{\Vert L\Vert^{2}}-\frac{2}{|S|^{2}(1-|S|^{2})}Z\right).
\end{align}
Since one can reconstruct the flux via
\begin{equation}
F=\frac{1}{\Vert L\Vert^{2}}\bigl(L\wedge\imath_{L}F+\star(L\wedge\imath_{L}\star F)\bigr),
\end{equation}
and the electric flux $f$ is determined by \eqref{eq:f_su2}, the two-form conditions completely fix the magnetic flux (and thus also the curvature of the connection one-form in \eqref{eq:connection}). As $|S|=\ee^{-3\Delta}\rho$ and $\rho$ is a coordinate on $X$, the above expressions can be thought of as depending on the function $\Delta$ and the torsion of the $\SU 2$ structure via the derivatives of $j_{\SU 2}$ and $\im\theta_{\SU 2}$.

Finally, let us consider the three-form conditions \eqref{eq:varphi-1}--\eqref{eq:Omega}. Since there are no independent invariant three-forms, we expect these conditions will reduce to conditions on the one-forms and $\SU 2$ structure forms, or to give consistency conditions for the flux. We begin by noting that $\varphi$ and $\phi$ are uncharged under $L$, whereas $\Omega$ is charge $-4m$. Taking $\imath_{L}$ of the conditions and rewriting them in terms of $\mathcal{L}_{L}$, one finds that the $\phi$ and $\Omega$ conditions exactly reproduce \eqref{eq:2-form_su2}, while the $\varphi$ condition fixes the derivative of $\re\theta_{\SU 2}$ as
\begin{equation}
\ee^{-3\Delta}\dd\bigl(\ee^{3\Delta}(1-|S|^{2})\re\theta_{\SU 2}\bigr)=-8m\imath_{L}\imath_{K}\star\varphi-2\imath_{L}(J\cdot F).
\end{equation}
At this point, one should check the remaining differential and algebraic conditions for any further constraints on the structure. In particular, since $S\neq0$, the Bianchi identities should be implied by the supersymmetry conditions. We leave the analysis of the full system to future work.

\section{Conclusions and outlook\label{sec:Conclusions-and-outlook}}

In this paper, we have presented the supersymmetry conditions for $N=(2,0)$ AdS$_{3}$ backgrounds of eleven-dimensional supergravity. We used the formalism of local $G$-structures to convert the Killing spinor equations into differential and algebraic conditions on a set of differential forms defined as bilinears of the Killing spinors. We paid special attention to a distinguished Killing vector field which is nowhere-vanishing for AdS solutions, and argued that it captures the R-symmetry of the dual $(2,0)$ SCFT. Using our results, we recovered all known classes of supersymmetric AdS$_{3}$ solutions and found two new classes of Minkowski solution: one with electric flux describing M5-branes wrapping SLAG three-cycles in eight-manifolds with an $\SU 3$ structure, and another describing M5-branes wrapping a Kähler two-cycle and a Killing direction. We also discussed some features of the cases of a generic local $\SU 3$ structure and a certain local $\SU 2$ structure. Both of these cases deserve further analysis similar to that of \cite{1207.3082}.

We considered the differential and algebraic constraints implied by supersymmetry in the supergravity limit of M-theory. This is not the end of the story; supergravity solutions can lift to string or M-theory backgrounds only if the fluxes satisfy certain quantisation conditions. Furthermore, backgrounds will generically have a tadpole, giving rise to a spacetime potential which will destabilise the vacuum. Together, these impose further constraints on the background. For example, if the eight-manifold admits a nowhere-vanishing spinor of each chirality (a $\Gx 2$ structure), the Euler number of the internal space vanishes and so $X_{8}$, the first correction to the $G$-flux equation of motion~\cite{hep-th/9908088}, must integrate to zero. This then restricts us to $\int_{X}G\wedge G=0$ solutions (though $X_{8}$ itself does not have to vanish and so can still allow for compact solutions to the equations of motion). In the case without nowhere-vanishing spinors of definite chirality, such as the local $\SU 2$ or $\SU 3$ cases we considered, one no longer has the usual relation between $X_{8}$ and the Euler number. Clearly, this deserves further investigation.

We did not give a thorough analysis of the case where $X$ admits a local $\SU 2$ structure, however it is easy to imagine how one would generate such a background. Starting with an AdS$_{3}$ solution where $(K,\re P,\im P)$ are Killing and generate a $\Uni 1^{3}$ isometry (in addition to the R-symmetry), one performs a $\gamma$-deformation à la Lunin and Maldacena. The resulting geometry should preserve $(2,0)$ supersymmetry, since the R-symmetry is untouched. Similarly, it would be interesting to check whether the $(2,0)$ solutions generated by non-abelian T-duality in \cite{Bea:2015fja,Araujo:2015npa} admit local $\SU 2$ structures.

Similarly to their type IIB cousins, it would be interesting to understand the geometric dual of $c$-extremisation~\cite{1211.4030,1302.4451} for the backgrounds we have considered. There is already an analysis of AdS$_{3}$ backgrounds~\cite{Couzens:2022agr} that give the massive IIA version of GK geometries~\cite{Kim:2005ez,Kim:2006qu,Gauntlett:2007ts}, where one can compute the central charge using only the topology of the solution. There is no obstruction to setting the Romans mass to zero in this analysis, and so this class of solution should lift to M-theory. More generally, for AdS$_{d}$ backgrounds with $d\geq4$, generalised geometry provides a guide for the gravity version of the various extremisation procedures that determine the R-symmetry in the dual field theories~\cite{1011.4296,Ashmore:2016qvs}. For example, the R-symmetry for AdS$_{4}$ flux backgrounds should be fixed by extremising a certain quartic functional, while for AdS$_{5}$ it is a cubic functional, generalising the well-known $\mathcal{F}$- and $a$-maximisation principles respectively~\cite{hep-th/0304128,1012.3210,1103.1181}. Unfortunately, AdS$_{3}$ backgrounds do not seem to admit a useful generalised geometric description along the lines of \cite{Coimbra:2014uxa,Coimbra:2017fqv}, and so it is at present unclear how $c$-extremisation should be phrased for generic flux backgrounds.\footnote{Though see, for example, \cite{Galli:2022idq} for recent work on using $\Ex{8(8)}$ exceptional field theory to construct consistent truncations to three-dimensional supergravity with AdS$_{3}$ vacua for varying amounts of supersymmetry.} The results in this paper will hopefully provide a starting point for this problem.

Finally, it would be interesting to study the superpotential that governs the three-dimensional effective theories that one would find after compactification. This has been investigated for AdS$_{3}$ backgrounds in type IIA and heterotic supergravity~\cite{2005.05246,deIaOssa:2019cci}, and it seems likely that there are a variety of questions that one could ask in M-theory~\cite{hep-th/0201227}. For example, using gauged supergravity, the superpotential is known to encode the central charge and R-symmetry of the dual SCFT for certain classes of $(2,0)$ backgrounds that come from wrapped M5-branes~\cite{Karndumri:2015sia}. It would be interesting to extend this result by analysing the superpotential that captures supersymmetry for a general background. Indeed, this sort of analysis would be essential for understanding whether there is a KKLT-like landscape of AdS$_{3}$ vacua with small cosmological constants in M-theory. We hope to return to this question in future work.

\subsection*{Acknowledgements}

It is a pleasure to thank André Coimbra, Charles Strickland-Constable and Eirik Eik Svanes for useful discussions, and Savdeep Sethi for collaboration on a related project that spurred this work. The author is supported in part by NSF Grant No.~PHY2014195 and in part by the Kadanoff Center for Theoretical Physics. The author also acknowledges the support of the European Union’s Horizon 2020 research and innovation program under the Marie Sk\l{}odowska-Curie grant agreement No.~838776.

\appendix

\section{Conventions\label{app:conventions}}

\subsection{Geometry}

We define the Hodge star on the eight-manifold $X$ as
\begin{equation}
\alpha\lrcorner\beta\vol_{8}=\star\alpha\wedge\beta,
\end{equation}
where $\alpha$ and $\beta$ are $p$-forms. With this convention, it obeys $\star^{2}\alpha=(-1)^{p}\alpha.$ Explicitly, in components, the Hodge star acts as
\begin{equation}
(\star\alpha)_{a_{1}\dots a_{8-p}}=\frac{1}{p!}\epsilon_{a_{1}\dots a_{8-p}b_{1}\dots b_{p}}\alpha^{b_{1}\dots b_{p}}.
\end{equation}
We sometimes denote by $\alpha^{\sharp}$ the $p$-vector given by raising the indices of $\alpha$ using the metric on $X$, though often leave it implicit when context is sufficient to distinguish between them.

We denote the contraction of a $p$-vector $v$ into a $q$-form $\alpha$ by $v\lrcorner\alpha$ or $\imath_{v}\alpha$. Given a frame $\{\hat{e}_{a}\}$ for $TX$ and a coframe $\{e^{a}\}$ for $T^{*}X$, the contraction is given in components by
\begin{equation}
(v\lrcorner\alpha)_{a_{1}\dots a_{q-p}}=\frac{1}{p!}v^{b_{1}\dots b_{p}}\alpha_{b_{1}\dots b_{p}a_{1}\dots a_{q-p}}.
\end{equation}
We also denote the norm of a $p$-form $\alpha$ by
\begin{equation}
\Vert\alpha\Vert^{2}\equiv\alpha\lrcorner\alpha.
\end{equation}
For a two-form $r$ and a $p$-form $\alpha$, we define $r\cdot\alpha$ to be the $\mathfrak{gl}_{d}$ action of $r$ as an endomorphism on $\alpha$. In components, this is given by
\begin{equation}
(r\cdot\alpha)_{a_{1}\dots a_{p}}=-r^{b}{}_{a_{1}}\alpha_{ba_{2}\dots a_{p}}-\dots-r^{b}{}_{a_{p}}\alpha_{a_{1}\dots a_{p-1}b}.
\end{equation}

\subsection{Clifford algebra and spinors}

A useful review of gamma matrices in any dimension can be found in \cite{1905.00429}. Some useful commutator and anticommutator identities can be found in \cite{0711.1436}. The relevant Clifford algebra in eleven dimensions is $\op{Cliff}(1,10)$, which we take to be generated by $\Gamma^{M}$ such that
\begin{equation}
\Gamma_{11}=\Gamma_{0}\dots\Gamma_{10}=-\id_{32}.
\end{equation}

\paragraph{Three dimensions}

An explicit representation of the Clifford algebra in three dimensions is given by the Pauli matrices as
\begin{equation}
\gamma_{0}=\ii\sigma_{1},\qquad\gamma_{1}=\sigma_{2},\qquad\gamma_{2}=\sigma_{3}.
\end{equation}
Note that $\gamma_{0}$ is anti-hermitian while $\gamma_{1}$ and $\gamma_{2}$ are hermitian, in agreement with
\begin{equation}
\{\gamma_{\mu},\gamma_{\nu}\}=2\eta_{\mu\nu},\qquad\eta_{\mu\nu}=\op{diag}(-1,1,1).
\end{equation}
One can check that
\begin{equation}
\gamma_{\mu}=\tfrac{1}{2}\epsilon_{\mu\nu\rho}\gamma^{\nu\rho},\qquad\gamma_{\mu\nu}=-\epsilon_{\mu\nu\rho}\gamma^{\rho},
\end{equation}
where the Levi-Civita symbol in Minkowski space obeys $\epsilon_{012}=-\epsilon^{012}$.

\paragraph{Eight dimensions}

We choose the gamma matrices in eight dimensions to be $16\times16$ real symmetric matrices with $\gamma_{9}^{2}=\id_{16}$, which satisfy
\begin{equation}
\{\gamma_{m},\gamma_{n}\}=2\delta_{mn},\qquad\gamma_{9}=\gamma_{1}\dots\gamma_{8}.
\end{equation}
An explicit realisation, useful for calculations, is given by
\begin{align}
\begin{split}\gamma_{1} & =\sigma_{1}\otimes\id_{2}\otimes\id_{2}\otimes\id_{2},\\
\gamma_{3} & =\sigma_{2}\otimes\sigma_{2}\otimes\sigma_{1}\otimes\id_{2},\\
\gamma_{5} & =\sigma_{2}\otimes\sigma_{1}\otimes\id_{2}\otimes\sigma_{2},\\
\gamma_{7} & =\sigma_{2}\otimes\id_{2}\otimes\sigma_{2}\otimes\sigma_{1},
\end{split}
 & \begin{split}\gamma_{2} & =\sigma_{3}\otimes\id_{2}\otimes\id_{2}\otimes\id_{2},\\
\gamma_{4} & =\sigma_{2}\otimes\sigma_{2}\otimes\sigma_{3}\otimes\id_{2},\\
\gamma_{6} & =\sigma_{2}\otimes\sigma_{3}\otimes\id_{2}\otimes\sigma_{2},\\
\gamma_{8} & =\sigma_{2}\otimes\id_{2}\otimes\sigma_{2}\otimes\sigma_{3},
\end{split}
\label{eq:gamma_realisation}
\end{align}
where the $\sigma_{i}$ are the Pauli matrices and $\id_{2}$ is the $2\times2$ identity matrix. With this choice, the highest-rank gamma matrix is given by $\gamma_{9}=\sigma_{2}\otimes\sigma_{2}\otimes\sigma_{2}\otimes\sigma_{2}$. All intertwiners are trivial for this realisation.\footnote{This is the case $\eta=\epsilon=-1$ in the notation of \cite{1905.00429}.} In particular, this means that one has $\xi^{\charge}=\xi^{*}$ and $\bar{\xi}=\xi^{\dagger}$ for a spinor $\xi$ in eight dimensions.

\subsection{Gamma matrices and spinor bilinears in eight dimensions\label{sec:Gamma-matrices-in}}

We collect here some useful properties of gamma matrices in eight dimensions. Compared with \cite{1905.00429}, this corresponds to $(\epsilon,\eta)=(-1,-1)$ with $d=s=8$ and $t=0$. Since $t=0$, the $A$ intertwiner is trivial, $A=\id_{16}$. Next, the $C$ intertwiner obeys
\begin{equation}
\gamma_{m}^{\transpose}=C\gamma_{m}C^{-1},\qquad C^{\transpose}=C.
\end{equation}
One can thus take $C=\id_{16}$ so that the gamma matrices are real and symmetric. The $B$ intertwiner, defined as $(A^{-1}C)^{\transpose}$ is then also trivial. This means we have $\bar{\xi}=\xi^{\dagger}$, $\xi^{\text{c}}=\xi^{*}$, and $\bar{\xi}^{\text{c}}=\xi^{\transpose}$ for any spinor $\xi$, as we stated in the previous subsection. With this choice, the rank-$n$ gamma matrices have the following symmetry properties:
\begin{equation}
\gamma_{(n)}=\begin{cases}
\text{symmetric} & n=0,1,4,5,8,\\
\text{antisymmetric} & n=2,3,6,7.
\end{cases}
\end{equation}
We also have the following (anti)commutation properties with $\gamma_{9}$:
\begin{equation}
\gamma_{9}\gamma_{(n)}=\begin{cases}
\gamma_{(n)}\gamma_{9} & n=\text{even},\\
-\gamma_{(n)}\gamma_{9} & n=\text{odd},
\end{cases}
\end{equation}
so that, for example, $\gamma_{9}$ anticommutes with rank-one gamma matrices. Using this one can constrain which bilinears are non-vanishing. 

First, for any spinors $\xi_{1}$ and $\xi_{2}$, we have
\begin{align}
\begin{split}(\bar{\xi}_{1}\gamma_{(n)}\xi_{2})^{*} & =(-1)^{n(n-1)/2}\bar{\xi}_{2}\gamma_{(n)}\xi_{1}\\
 & =\begin{cases}
+\bar{\xi}_{2}\gamma_{(n)}\xi_{1} & n=0,1,4,5,8,\\
-\bar{\xi}_{2}\gamma_{(n)}\xi_{1} & n=2,3,6,7,
\end{cases}
\end{split}
\\
\begin{split}(\bar{\xi}_{1}\gamma_{9}\gamma_{(n)}\xi_{2})^{*} & =(-1)^{n(n+1)/2}\bar{\xi}_{2}\gamma_{9}\gamma_{(n)}\xi_{1}\\
 & =\begin{cases}
+\bar{\xi}_{2}\gamma_{9}\gamma_{(n)}\xi_{1} & n=0,3,4,7,8,\\
-\bar{\xi}_{2}\gamma_{9}\gamma_{(n)}\xi_{1} & n=1,2,5,6.
\end{cases}
\end{split}
\end{align}
From this it follows that
\begin{align}
\bar{\xi}\gamma_{(n)}\xi & =\begin{cases}
\text{real} & n=0,1,4,5,8,\\
\text{imaginary} & n=2,3,6,7,
\end{cases}\\
\bar{\xi}\gamma_{9}\gamma_{(n)}\xi & =\begin{cases}
\text{real} & n=0,3,4,7,8,\\
\text{imaginary} & n=1,2,5,6.
\end{cases}
\end{align}
For charge conjugate bilinears, one has
\begin{align}
\begin{split}\bar{\xi}_{1}^{\text{c}}\gamma_{(n)}\xi_{2} & =(-1)^{n(n-1)/2}\bar{\xi}_{2}^{\text{c}}\gamma_{(n)}\xi_{1}\\
 & =\begin{cases}
+\bar{\xi}_{2}^{\text{c}}\gamma_{(n)}\xi_{1} & n=0,1,4,5,8,\\
-\bar{\xi}_{2}^{\text{c}}\gamma_{(n)}\xi_{1} & n=2,3,6,7,
\end{cases}
\end{split}
\\
\begin{split}\bar{\xi}_{1}^{\text{c}}\gamma_{9}\gamma_{(n)}\xi_{2} & =(-1)^{n(n+1)/2}\bar{\xi}_{2}^{\text{c}}\gamma_{9}\gamma_{(n)}\xi_{1}\\
 & =\begin{cases}
+\bar{\xi}_{2}^{\text{c}}\gamma_{9}\gamma_{(n)}\xi_{1} & n=0,3,4,7,8,\\
-\bar{\xi}_{2}^{\text{c}}\gamma_{9}\gamma_{(n)}\xi_{1} & n=1,2,5,6.
\end{cases}
\end{split}
\end{align}
In particular, this implies the following vanishing results:
\begin{align}
\bar{\xi}^{\text{c}}\gamma_{(n)}\xi & =0\quad\text{if}\quad n=2,3,6,7,\\
\bar{\xi}^{\text{c}}\gamma_{9}\gamma_{(n)}\xi & =0\quad\text{if}\quad n=1,2,5,6.
\end{align}
Finally, for Weyl spinors $\xi_{1,2}$, one observes that spinors of opposite chirality are orthogonal and that odd-rank gamma matrices flip the chirality of a Weyl spinor, which means
\begin{equation}
\bar{\xi}_{2}\gamma_{(n)}\xi_{1}=0\quad\text{if }\quad\begin{cases}
\text{\ensuremath{\xi_{1}} and \ensuremath{\xi_{2}} are same chirality and \ensuremath{n=\text{odd},}}\\
\text{\ensuremath{\xi_{1}} and \ensuremath{\xi_{2}} are opposite chirality and \ensuremath{n=\text{even.}}}
\end{cases}
\end{equation}

\section{Algebraic Killing spinor equations\label{sec:Algebraic-Killing-spinor}}

Here we give the algebraic constraints that one derives from requiring that the external component of the gravitino variation \eqref{eq:KS_eq_alg} vanishes. These conditions first appeared in \cite{1212.6918,1311.5901}, where they were derived by lifting the Killing spinors to a nine-dimensional auxiliary space. In our conventions, the zero- to four-form conditions are as follows.

\paragraph{Scalar conditions}

\begin{align}
0 & =-2m\zeta+K\lrcorner\dd\Delta-\tfrac{1}{6}\Phi\lrcorner F,\\
0 & =L\lrcorner f,\\
0 & =-2mS+P\lrcorner\dd\Delta-\tfrac{1}{6}\Gamma\lrcorner F,\\
0 & =-2m+\tfrac{1}{3}K\lrcorner f-\tfrac{1}{6}(\star\Phi)\lrcorner F,\\
0 & =L\lrcorner\dd\Delta,\\
0 & =\tfrac{1}{3}P\lrcorner f-\tfrac{1}{6}(\star\Gamma)\lrcorner F.
\end{align}

\paragraph{One-form conditions}

\begin{align}
0 & =\dd\Delta-\tfrac{1}{3}\zeta f+\tfrac{1}{6}F\lrcorner(\star\phi),\\
0 & =2mL-\dd\Delta\lrcorner J+\tfrac{1}{3}f\lrcorner\omega+\tfrac{1}{6}\varphi\lrcorner F,\\
0 & =-\tfrac{1}{3}Sf+\tfrac{1}{6}F\lrcorner(\star\Omega),\\
0 & =-\dd\Delta\lrcorner\omega+\tfrac{1}{3}f\lrcorner J+\tfrac{1}{6}F\lrcorner(\star\varphi),\\
0 & =2mK+\zeta\,\dd\Delta-\tfrac{1}{3}f+\tfrac{1}{6}\phi\lrcorner F,\\
0 & =2mP+S\,\dd\Delta+\tfrac{1}{6}\Omega\lrcorner F.
\end{align}

\paragraph{Two-form conditions}

\begin{align}
0 & =-2m\omega+\dd\Delta\lrcorner\varphi-\tfrac{1}{3}f\wedge L+\tfrac{1}{6}J\lrcorner F+\tfrac{1}{6}\omega\lrcorner\star F,\\
0 & =-\dd\Delta\wedge K+\tfrac{1}{3}f\lrcorner\phi-\tfrac{1}{36}\Phi_{a}{}^{bcd}F_{bcde}e^{ea},\\
0 & =-\dd\Delta\wedge P+\tfrac{1}{3}f\lrcorner\Omega-\tfrac{1}{36}\Gamma_{a}{}^{bcd}F_{bcde}e^{ea},\\
0 & =-2mJ-\dd\Delta\wedge L+\tfrac{1}{3}f\lrcorner\varphi+\tfrac{1}{6}\omega\lrcorner F+\tfrac{1}{6}J\lrcorner\star F,\\
0 & =\dd\Delta\lrcorner\phi-\tfrac{1}{3}f\wedge K-\tfrac{1}{36}(\star\Phi)_{a}{}^{abcd}F_{bcde}e^{ea},\\
0 & =\dd\Delta\lrcorner\Omega-\tfrac{1}{3}f\wedge P-\tfrac{1}{36}(\star\Gamma)_{a}{}^{bcd}F_{bcde}e^{ea}.
\end{align}

\paragraph{Three-form conditions}

\begin{align}
0 & =3\dd\Delta\wedge J-f\wedge\omega+\tfrac{1}{8}\varphi_{a}{}^{bc}F_{bcde}e^{dea}+\tfrac{1}{2}L\lrcorner\star F,\\
0 & =2m\phi-\dd\Delta\lrcorner\Phi+\tfrac{1}{3}f\lrcorner\star\Phi-\tfrac{1}{6}K\lrcorner F-\tfrac{1}{72}(\star\phi)_{ab}{}^{cde}F_{cdef}e^{fab},\\
0 & =2m\Omega-\dd\Delta\lrcorner\Gamma+\tfrac{1}{3}f\lrcorner\star\Gamma-\tfrac{1}{6}P\lrcorner F-\tfrac{1}{72}(\star\Omega)_{ab}{}^{cde}F_{cdef}e^{fab},\\
0 & =-\dd\Delta\lrcorner\Phi+\tfrac{1}{3}f\lrcorner\Phi+\tfrac{1}{24}\phi_{a}{}^{bc}F_{bcde}e^{dea}+\tfrac{1}{6}K\lrcorner\star F,\\
0 & =2m\varphi+3\dd\Delta\wedge\omega-f\wedge J-\tfrac{1}{6}L\lrcorner F-\tfrac{1}{72}(\star\varphi)_{ab}{}^{cde}F_{cdef}e^{fab},\\
0 & =-\dd\Delta\lrcorner\star\Gamma+\tfrac{1}{3}f\lrcorner\Gamma+\tfrac{1}{24}\Omega_{a}{}^{bc}F_{bcde}e^{dea}+\tfrac{1}{6}P\lrcorner\star F.
\end{align}

\paragraph{Four-form conditions}

\begin{align}
0 & =-\dd\Delta\wedge\varphi-\tfrac{1}{3}f\lrcorner\star\varphi+\tfrac{1}{36}J_{a}{}^{b}F_{bcde}e^{cdea}+\tfrac{1}{216}(\star\omega)_{abc}{}^{def}F_{defg}e^{gabc},\\
0 & =-2m\Phi-\dd\Delta\wedge\phi-\tfrac{1}{3}f\lrcorner\star\phi-\tfrac{1}{6}\zeta\,F-\tfrac{1}{6}\star F+\tfrac{1}{48}(\star\Phi)_{ab}{}^{cd}F_{cdef}e^{efab},\\
0 & =-2m\Gamma-\dd\Delta\wedge\Omega-\tfrac{1}{3}f\lrcorner\star\Omega-\tfrac{1}{6}S\,F+\tfrac{1}{48}(\star\Gamma)_{ab}{}^{cd}F_{cdef}e^{efab}.
\end{align}

\section{An orthonormal frame for the local \texorpdfstring{$\SU3$}{SU(3)} structure\label{sec:An-orthonormal-frame}}

In this appendix, we relate the local $\SU 3$ structure specified by the pair of Majorana Killing spinors to a canonical $\SU 3$ structure on $X$.

\subsection{A canonical \texorpdfstring{$\SU3$}{SU(3)} structure in eight dimensions\label{subsec:A-canonical-}}

Following a similar strategy in \cite[Appendix C]{Gauntlett:2004zh}, the idea is to first define a pair of canonical $\SU 4$ structures, or equivalently a pair of orthogonal complex chiral spinors, and then relate these to the spinors that solve the Killing spinor equations.

An $\SU 4$ structure on an eight-manifold is defined by two invariant real orthogonal spinors of the same chirality in eight dimensions~\cite{Gauntlett:2003cy}. We will need two of these $\SU 4$ structures. The first $\SU 4$ structure is defined by the pair $(\epsilon_{1},\epsilon_{3})$ such that
\begin{equation}
\begin{aligned}\gamma_{1234}\epsilon_{1} & =\gamma_{5678}\epsilon_{1}=\gamma_{1256}\epsilon_{1}=-\gamma_{1357}\epsilon_{1}=-\epsilon_{1},\\
\gamma_{1234}\epsilon_{3} & =\gamma_{5678}\epsilon_{3}=\gamma_{1256}\epsilon_{3}=+\gamma_{1357}\epsilon_{3}=-\epsilon_{3}.
\end{aligned}
\end{equation}
These conditions imply that both spinors are positive chirality
\begin{equation}
\gamma_{9}\epsilon_{1}=+\epsilon_{1},\qquad\gamma_{9}\epsilon_{3}=+\epsilon_{3}.
\end{equation}
Upon imposing $\gamma_{78}\epsilon_{1}=-\epsilon_{3}$ and normalising the spinors to one,\footnote{Recall that $\bar{\epsilon}_{i}=\epsilon_{i}^{\transpose}$ since Majorana spinors are real with our choice of gamma matrices.}
\begin{equation}
\bar{\epsilon}_{1}\epsilon_{1}=\bar{\epsilon}_{3}\epsilon_{3}=1,
\end{equation}
we recover an $\SU 4$ structure in the usual orthonormal frame. Explicitly, we define the complex Weyl spinor
\begin{equation}
\eta_{1}=\frac{1}{\sqrt{2}}(\epsilon_{1}+\ii\epsilon_{3}).
\end{equation}
The first $\SU 4$ structure is then given by
\begin{equation}
\begin{aligned}-\ii\bar{\eta}_{1}\gamma_{(2)}\eta_{1} & =e^{12}+e^{34}+e^{56}+e^{78},\\
\eta_{1}\gamma_{(4)}\eta_{1} & =(e^{1}+\ii e^{2})\wedge(e^{3}+\ii e^{4})\wedge(e^{5}+\ii e^{6})\wedge(e^{7}+\ii e^{8}).
\end{aligned}
\end{equation}

The second $\SU 4$ structure is defined by the pair $(\epsilon_{2},\epsilon_{4})$ such that
\begin{equation}
\begin{aligned}\gamma_{1234}\epsilon_{2} & =-\gamma_{5678}\epsilon_{2}=\gamma_{1256}\epsilon_{2}=-\gamma_{1357}\epsilon_{2}=-\epsilon_{2},\\
\gamma_{1234}\epsilon_{4} & =-\gamma_{5678}\epsilon_{4}=\gamma_{1256}\epsilon_{4}=+\gamma_{1357}\epsilon_{4}=-\epsilon_{4}.
\end{aligned}
\end{equation}
These imply that both spinors are negative chirality so that
\begin{equation}
\gamma_{9}\epsilon_{2}=-\epsilon_{2},\qquad\gamma_{9}\epsilon_{4}=-\epsilon_{4}.
\end{equation}
Similarly to above, we also impose
\begin{equation}
\gamma_{12}\epsilon_{2}=-\epsilon_{4},\qquad\bar{\epsilon}_{2}\epsilon_{2}=\bar{\epsilon}_{4}\epsilon_{4}=1.
\end{equation}
We then define the complex Weyl spinor
\begin{equation}
\eta_{2}=\frac{1}{\sqrt{2}}(\epsilon_{2}+\ii\epsilon_{4}),
\end{equation}
so that the second $\SU 4$ structure is given by
\begin{equation}
\begin{aligned}-\ii\bar{\eta}_{2}\gamma_{(2)}\eta_{2} & =e^{12}+e^{34}+e^{56}-e^{78},\\
\eta_{2}\gamma_{(4)}\eta_{2} & =(e^{1}+\ii e^{2})\wedge(e^{3}+\ii e^{4})\wedge(e^{5}+\ii e^{6})\wedge(e^{7}-\ii e^{8}).
\end{aligned}
\end{equation}
Note that the unit-norm spinors can be written in terms of $\epsilon_{1}$ as
\begin{equation}
\epsilon_{2}=\gamma_{8}\epsilon_{1},\qquad\epsilon_{3}=-\gamma_{78}\epsilon_{1},\qquad\epsilon_{4}=\gamma_{7}\epsilon_{1}.\label{eq:su3_spinors}
\end{equation}

The $\SU 3$ structure on which the two $\SU 4$'s intersect is picked out by the complex one-form
\begin{equation}
\ii\bar{\eta}_{2}\gamma_{(1)}\eta_{1}=e^{7}+\ii e^{8}.
\end{equation}
With this, we can write the $\SU 4$ structures as
\begin{equation}
\begin{aligned}-\ii\bar{\eta}_{1}\gamma_{(2)}\eta_{1} & =j+e^{78}, & -\ii\bar{\eta}_{2}\gamma_{(2)}\eta_{2} & =j-e^{78},\\
\eta_{1}\gamma_{(4)}\eta_{1} & =\theta\wedge(e^{7}+\ii e^{8}), & \eta_{2}\gamma_{(4)}\eta_{2} & =\theta\wedge(e^{7}-\ii e^{8}),
\end{aligned}
\end{equation}
where the common $\SU 3$ structure is given by
\begin{equation}
k=e^{7}+\ii e^{8},\qquad j=e^{12}+e^{34}+e^{56},\qquad\theta=(e^{1}+\ii e^{2})\wedge(e^{3}+\ii e^{4})\wedge(e^{5}+\ii e^{6}).
\end{equation}
This is the canonical $\SU 3$ structure that we use for calculations in the main text. With our conventions, these forms obey
\begin{equation}
\begin{gathered}\tfrac{1}{6}j^{3}\wedge e^{78}=\tfrac{\ii}{8}\theta\wedge\bar{\theta}\wedge e^{78}=\star1,\qquad\tfrac{1}{6}j^{3}=\tfrac{\ii}{8}\theta\wedge\bar{\theta},\\
\star j=\tfrac{1}{2}j^{2}\wedge e^{78},\qquad\tfrac{1}{2}\star j^{2}=j\wedge e^{78},\qquad\tfrac{1}{6}\star j^{3}=e^{78},\\
\star\theta=\ii\theta\wedge e^{78},\qquad\star e^{7}=-\tfrac{1}{6}j^{3}\wedge e^{8},\qquad\star e^{8}=\tfrac{1}{6}j^{3}\wedge e^{7}.
\end{gathered}
\end{equation}

Alternatively, the spinors can be thought of as defining a pair of canonical $\Gx 2$ structures which intersect on an $\SU 3$. Using the above unit-norm spinors, we define the non-chiral Majorana spinors
\begin{equation}
\theta_{1}=\frac{1}{\sqrt{2}}(\epsilon_{1}+\epsilon_{2}),\qquad\theta_{2}=\frac{1}{\sqrt{2}}(\epsilon_{3}+\epsilon_{4}).
\end{equation}
In terms of these, the two $\Gx 2$ structures in eight dimensions are defined by
\begin{equation}
\begin{aligned}\theta_{1}\gamma_{(1)}\theta_{1} & =e^{8}, & \theta_{2}\gamma_{(1)}\theta_{2} & =e^{8},\\
\theta_{1}\gamma_{(4)}\theta_{1} & =\re\theta\wedge e^{7}-\tfrac{1}{2}j^{2}, & \theta_{2}\gamma_{(4)}\theta_{2} & =-\re\theta\wedge e^{7}-\tfrac{1}{2}j^{2},\\
\theta_{1}\gamma_{(3)}\gamma_{9}\theta_{1} & =\im\theta+j\wedge e^{7}, & \theta_{2}\gamma_{(3)}\gamma_{9}\theta_{2} & =-\im\theta+j\wedge e^{7}.
\end{aligned}
\end{equation}
The common $\SU 3$ structure is then picked out by the real one-form
\begin{equation}
\theta_{2}\gamma_{(1)}\gamma_{9}\theta_{1}=e^{7}.
\end{equation}

\subsection{A local \texorpdfstring{$\SU3$}{SU(3)} structure\label{subsec:A-local-}}

We now want to relate the spinors that define the canonical $\SU 3$ structure above to the pair of non-chiral Majorana spinors $\chi_{i}$ that appear in the Killing spinor equations. Without loss of generality, we can write these in terms of the spinors which define the canonical $\SU 3$ structure as
\begin{equation}
\begin{aligned}\chi_{1} & =\tfrac{1}{\sqrt{2}}(a_{1}\epsilon_{1}+a_{2}\epsilon_{2}),\\
\chi_{2} & =\tfrac{1}{\sqrt{4}}(b_{1}\epsilon_{1}+b_{2}\epsilon_{2}+b_{3}\epsilon_{3}+b_{4}\epsilon_{4}),
\end{aligned}
\label{eq:KS_coeffs}
\end{equation}
where the coefficients $a_{i}$ and $b_{i}$ are real so that $\chi_{i}$ are Majorana.

First, one can solve the normalisation and orthogonality constraints for the spinors:
\begin{equation}
\begin{aligned}\chi_{1}\chi_{1} & \equiv1=\tfrac{1}{2}(a_{1}^{2}+a_{2}^{2}),\\
\chi_{2}\chi_{2} & \equiv1=\tfrac{1}{4}(b_{1}^{2}+b_{2}^{2}+b_{3}^{2}+b_{4}^{2}),\\
\chi_{2}\chi_{1} & \equiv0=\tfrac{1}{\sqrt{8}}(a_{1}b_{1}+a_{2}b_{2}).
\end{aligned}
\label{eq:norm_ortho}
\end{equation}
  One finds
\begin{equation}
a_{2}=m\sqrt{2-a_{1}^{2}},\qquad b_{1}=-mn\frac{1}{\sqrt{2}}\sqrt{2-a_{1}^{2}}\sqrt{4-b_{3}^{2}-b_{4}^{2}},\qquad b_{2}=n\frac{a_{1}}{\sqrt{2}}\sqrt{4-b_{3}^{2}-b_{4}^{2}},\label{eq:orth_norm_conds}
\end{equation}
where $m^{2}=n^{2}=1$ encode the allowed choices of signs. Note that we need
\begin{equation}
a_{1}^{2}\leq2,\qquad b_{3}^{2}+b_{4}^{2}\leq4,\label{eq:ineqality}
\end{equation}
for a solution to exist. We then match the remaining scalar bilinears with the parametrisation in terms of $\zeta$ and $S$ given in \eqref{eq:zeta_S_def}:
\begin{equation}
\begin{aligned}\zeta & =\tfrac{1}{2}(\chi_{1}\gamma_{9}\chi_{1}+\chi_{2}\gamma_{2}\chi_{2})\equiv\tfrac{1}{8}(2a_{1}^{2}-2a_{2}^{2}+b_{1}^{2}-b_{2}^{2}+b_{3}^{2}-b_{4}^{2}),\\
\re S & =\tfrac{1}{2}(\chi_{1}\gamma_{9}\chi_{1}-\chi_{2}\gamma_{2}\chi_{2})\equiv\tfrac{1}{8}(2a_{1}^{2}-2a_{2}^{2}-b_{1}^{2}+b_{2}^{2}-b_{3}^{2}+b_{4}^{2}),\\
\im S & =\chi_{2}\gamma_{9}\chi_{1}\equiv\tfrac{1}{\sqrt{8}}(a_{1}b_{1}-a_{2}b_{2}).
\end{aligned}
\end{equation}
Substituting in the solution in \eqref{eq:orth_norm_conds}, one finds these are given in terms of the coefficients in \eqref{eq:KS_coeffs} as
\begin{equation}
\begin{aligned}\zeta & =\tfrac{1}{8}\left(a_{1}^{2}(b_{3}^{2}+b_{4}^{2})-2b_{4}^{2}\right),\\
\re S & =\tfrac{1}{8}\left(2(b_{4}^{2}-4)+a_{1}^{2}(8-b_{3}^{2}-b_{4}^{2})\right),\\
\im S & =-\tfrac{1}{2}mna_{1}\sqrt{2-a_{1}^{2}}\sqrt{4-b_{3}^{2}-b_{4}^{2}}.
\end{aligned}
\end{equation}
Using these expressions and the inequalities in \eqref{eq:ineqality}, it is then straightforward to show that
\begin{equation}
\zeta^{2}+|S|^{2}\leq1,
\end{equation}
which then allows the parametrisation of $\zeta$ and $S$ given in \eqref{eq:S_zeta_def}.

\subsection{Expressions for bilinears}

It is useful to give expressions for some of the bilinears that appear in Table \ref{tab:bilinears} in terms of the orthonormal frame defined by the canonical $\SU 3$ structure and the coefficients that appear in the Killing spinors in \eqref{eq:KS_coeffs}. Expressions for all of the bilinears can be found in the accompanying Mathematica notebook.

\paragraph{Scalars}

The scalar bilinears are
\begin{equation}
\bar{\chi}\chi=1,\qquad\bar{\chi}^{\text{c}}\chi=0,\qquad\bar{\chi}\gamma_{9}\chi=\zeta,\qquad\bar{\chi}^{\text{c}}\gamma_{9}\chi=S,
\end{equation}
where $\zeta$ is real and $S$ is complex. As we mentioned in Section \ref{sec:Analysis-of--structures}, the scalars bilinears are constrained by $\zeta^{2}+|S|^{2}\leq1.$

\paragraph{One-forms}

The one-form bilinears are

\begin{equation}
\bar{\chi}\gamma_{(1)}\chi=K,\qquad\bar{\chi}^{\text{c}}\gamma_{(1)}\chi=P,\qquad\bar{\chi}\gamma_{9}\gamma_{(1)}\chi=\ii L,\qquad\bar{\chi}^{\text{c}}\gamma_{9}\gamma_{(1)}\chi=0,
\end{equation}
where $K$ and $L$ are real, and $P$ is complex. Explicitly, the one-forms are given by
\begin{align}
4\,K & =(-b_{2}b_{3}+b_{1}b_{4})e^{7}+(2a_{1}a_{2}+b_{1}b_{2}+b_{3}b_{4})e^{8},\\
2\sqrt{2}\,L & =(a_{2}b_{3}+a_{1}b_{4})e^{7}+(-a_{2}b_{1}+a_{1}b_{2})e^{8},\\
4\re P & =(b_{2}b_{3}-b_{1}b_{4})e^{7}+(2a_{1}a_{2}-b_{1}b_{2}-b_{3}b_{4})e^{8},\\
2\sqrt{2}\im P & =(-a_{2}b_{3}+a_{1}b_{4})e^{7}+(a_{2}b_{1}+a_{1}b_{2})e^{8}.
\end{align}
Upon using the normalisation and orthogonality constraints in \eqref{eq:orth_norm_conds}, it is simple to check that the above satisfy the Fierz identities in \eqref{eq:fierz}. When $K$ and $L$ are linearly independent and $P$ is non-zero, the complex one-form $P$ can thus be written as a linear combination of $K$ and $L$, as is explicit above.

\paragraph{Two-forms and higher}

We do not find it instructive to give expressions for the two-forms and higher. Instead, explicit expressions for the remaining bilinears in Table \ref{tab:bilinears} are given in the accompanying Mathematica notebook.

\subsection{Some special cases}

In what remains of this appendix, we give specific choices for the coefficients in the spinor ansatz \eqref{eq:KS_coeffs} that reproduce the examples given in the main text.

\subsubsection{$L=0$\label{par:spinors L=00003D0}}

Combined with the constraints from the norms and orthogonality of the $\chi_{i}$ in \eqref{eq:norm_orthogonal}, it is simple to check that the solutions to $L=0$ are
\begin{equation}
a_{2}=m\sqrt{2-a_{1}^{2}},\quad b_{1}=0,\quad b_{2}=0,\quad b_{3}=n\sqrt{2}a_{1},\quad b_{4}=-\sqrt{2}mn\sqrt{2-a_{1}^{2}},
\end{equation}
where $|a_{1}|\leq\sqrt{2}$ and $m^{2}=n^{2}=1$. To match with the calculations in Section \ref{subsec:L=00003D0}, we fix the signs as $n=-m=1$ with $a_{1}=\sqrt{1+\zeta}$. These imply
\begin{equation}
S=0,\qquad L=K=0,\qquad P=\ii\sqrt{1-\zeta^{2}}(e^{7}+\ii e^{8}).\label{eq:S=00003D0 vectors-1}
\end{equation}

\subsubsection{$\zeta=0$ and $|S|=1$\label{subsec:Chiral-and-antichiral}}

Consider the case where the scalar bilinears satisfy
\begin{equation}
\zeta=0,\qquad S=\ee^{\ii\sigma}.
\end{equation}
Fixing some arbitrary signs, it is straightforward to see that the spinor coefficients can be taken to be
\begin{equation}
a_{1}=\sqrt{1+\cos\sigma},\quad a_{2}=-\sqrt{1-\cos\sigma},\quad b_{1}=\frac{\sqrt{2}\sin\sigma}{\sqrt{1+\cos\sigma}},\quad b_{2}=\frac{\sqrt{2}\sin\sigma}{\sqrt{1-\cos\sigma}},
\end{equation}
for $\sigma\in[0,\pi]$ with $b_{3}=b_{4}=0$. This choice agrees with the bilinears given in Section \ref{subsec:-holonomy-solutions}, and implies
\begin{equation}
\zeta=0,\qquad S=\ee^{\ii\sigma},\qquad K=0,\qquad-\ii\bar{S}P=L=e^{8}.
\end{equation}

\subsubsection{$S=0$\label{subsec:S=00003D0}}

With the constraints from the norms and orthogonality of the $\chi_{i}$ in \eqref{eq:norm_orthogonal}, the only solutions to $S=0$ are
\begin{equation}
a_{2}=m\sqrt{2-a_{1}^{2}},\quad b_{1}=0,\quad b_{2}=0,\quad b_{3}=n\sqrt{2}a_{1},\quad b_{4}=\sqrt{2}p\sqrt{2-a_{1}^{2}},
\end{equation}
where $|a_{1}|\leq\sqrt{2}$ and $m^{2}=n^{2}=p^{2}=1$. It is also straightforward to check that for $S=0$ we have
\begin{equation}
\begin{gathered}\zeta=a_{1}^{2}-1,\qquad\Vert L\Vert^{2}=\tfrac{1}{2}a_{1}^{2}(2-a_{1}^{2})(1+mnp),\\
\Vert K\Vert^{2}=\tfrac{1}{2}a_{1}^{2}(2-a_{1}^{2})(1+mnp),\qquad\Vert P\Vert^{2}=a_{1}^{2}(2-a_{1}^{2})(1-mnp).
\end{gathered}
\end{equation}
Taking $p=-mn$ returns us to the $L=0$ case of Section \ref{par:spinors L=00003D0}. For $L\neq0$, as in Section \ref{subsec:AdS-from-wrapping}, we take $p=mn$ with $m=n=1$, and $a_{1}=\sqrt{1+\zeta}$. This choice gives
\begin{equation}
S=0,\qquad K=\sqrt{1-\zeta^{2}}e_{8},\qquad L=\sqrt{1-\zeta^{2}}e_{7},\qquad P=0.
\end{equation}

\subsubsection{$\zeta=0$ and $|S|<1$ with $\beta=1$\label{subsec:-and}}

Consider the class of solutions with
\begin{equation}
\zeta=0,\qquad|S|\leq1,\qquad\beta=1.
\end{equation}
This gives the backgrounds considered in Section \ref{subsec:AdS-from-wrapping-1}. Fixing some arbitrary signs, the spinor coefficients can be taken to be
\begin{equation}
\begin{gathered}a_{1}=\sqrt{1+\re S},\quad a_{2}=\sqrt{1-\re S},\quad b_{1}=\frac{\sqrt{2}\im S}{\sqrt{1+\re S}},\quad b_{2}=-\frac{\sqrt{2}\im S}{\sqrt{1-\re S}},\\
b_{3}=-\frac{\sqrt{2}\sqrt{1-|S|^{2}}}{\sqrt{1+\re S}},\quad b_{4}=\frac{\sqrt{2}\sqrt{1-|S|^{2}}}{\sqrt{1-\re S}}.
\end{gathered}
\end{equation}
With this choice, one recovers the bilinears in \eqref{eq:ads_coass_bilinears}.

\subsubsection{$\zeta=0$ and $|S|<1$ with $\beta=|S|$\label{subsec:-and-1}}

Consider the class of solutions with
\begin{equation}
\zeta=0,\qquad|S|\leq1,\qquad\beta=|S|.
\end{equation}
This gives the backgrounds considered in Section \ref{subsec:New-example}. Fixing some arbitrary signs, the spinor coefficients can be taken to be
\begin{equation}
\begin{gathered}a_{1}=\sqrt{1+\re S},\quad a_{2}=\sqrt{1-\re S},\quad b_{1}=\frac{\sqrt{2}\im S}{\sqrt{1+\re S}},\quad b_{2}=-\frac{\sqrt{2}\im S}{\sqrt{1-\re S}},\\
b_{3}=\frac{\sqrt{2}\sqrt{1-|S|^{2}}}{\sqrt{1+\re S}},\quad b_{4}=\frac{\sqrt{2}\sqrt{1-|S|^{2}}}{\sqrt{1-\re S}}.
\end{gathered}
\end{equation}
This is the same as the previous case apart from a change of sign for $b_{3}$. As we show in Section \ref{subsec:New-example}, the supersymmetry conditions require that the phase of $S$ is constant on $X$, so we are free to take $S$ to be real. With this choice, one recovers the bilinears in \eqref{eq:kahler_2}.

\section{An orthonormal frame for the local \texorpdfstring{$\SU2$}{SU(2)} structure\label{sec:An-orthonormal-frame-1}}

In this appendix, we relate the local $\SU 2$ structure specified by the pair of Majorana Killing spinors to a canonical $\SU 2$ structure on $X$.

\subsection{A canonical \texorpdfstring{$\SU2$}{SU(2)} structure in eight dimensions}

Following the previous appendix, we define a pair of canonical $\SU 4$ structures which intersect on an $\SU 2$ rather than an $\SU 3$, and then relate these to the spinors that solve the Killing spinors equations.

We begin with the first $\SU 4$ defined by $(\epsilon_{1},\epsilon_{3}$). We take a positive chirality unit-norm spinor $\epsilon_{1}$ which is defined by the same set of projections as in the previous appendix:
\begin{equation}
\bar{\epsilon}_{1}\epsilon_{1}=1,\qquad\gamma_{1234}\epsilon_{1}=\gamma_{5678}\epsilon_{1}=\gamma_{1256}\epsilon_{1}=-\gamma_{1357}\epsilon_{1}=-\epsilon_{1}.
\end{equation}
As in the previous case in Appendix \ref{subsec:A-canonical-}, we define $\epsilon_{3}$ as
\begin{equation}
\epsilon_{3}=-\gamma_{78}\epsilon_{1}.
\end{equation}
The complex Weyl spinor
\begin{equation}
\eta_{1}=\frac{1}{\sqrt{2}}(\epsilon_{1}+\ii\epsilon_{3}),
\end{equation}
then defines an $\SU 4$ structure as
\begin{equation}
\begin{aligned}-\ii\bar{\eta}_{1}\gamma_{(2)}\eta_{1} & =e^{12}+e^{34}+e^{56}+e^{78},\\
\eta_{1}\gamma_{(4)}\eta_{1} & =(e^{1}+\ii e^{2})\wedge(e^{3}+\ii e^{4})\wedge(e^{5}+\ii e^{6})\wedge(e^{7}+\ii e^{8}).
\end{aligned}
\end{equation}

The second $\SU 4$ structure is defined by the pair $(\epsilon_{2},\epsilon_{6})$ such that
\begin{equation}
\epsilon_{2}=\gamma_{8}\epsilon_{1},\qquad\epsilon_{6}=\gamma_{5}\epsilon_{1}.
\end{equation}
This differs from \eqref{eq:su3_spinors} in the gamma matrix acting on the second spinor ($\gamma_{5}$ for $\epsilon_{6}$ instead of $\gamma_{7}$ for $\epsilon_{4}$). As we will see, this has the effect of picking out a common $\SU 2$ instead of $\SU 3$. The complex Weyl spinor
\begin{equation}
\eta_{2}=\frac{1}{\sqrt{2}}(\epsilon_{2}+\ii\epsilon_{6}),
\end{equation}
defines an $\SU 4$ structure given by
\begin{equation}
\begin{aligned}-\ii\bar{\eta}_{2}\gamma_{(2)}\eta_{2} & =e^{14}+e^{23}-e^{58}+e^{67},\\
\eta_{2}\gamma_{(4)}\eta_{2} & =(e^{1}+\ii e^{4})\wedge(e^{2}+\ii e^{3})\wedge(-e^{5}+\ii e^{8})\wedge(e^{6}+\ii e^{7}).
\end{aligned}
\end{equation}

The $\SU 2$ structure on which the two $\SU 4$'s intersect is picked out by the pair of complex one-forms
\begin{align}
\ii\bar{\eta}_{2}\gamma_{(1)}\eta_{1} & =\tfrac{1}{2}(e^{5}+\ii e^{6}+e^{7}+\ii e^{8}),\\
\ii\eta_{2}\gamma_{(1)}\eta_{1} & =\tfrac{1}{2}(-e^{5}-\ii e^{6}+e^{7}+\ii e^{8}).
\end{align}
With this, we can write the $\SU 4$ structures as
\begin{equation}
\begin{aligned}-\ii\bar{\eta}_{1}\gamma_{(2)}\eta_{1} & =j_{3}+e^{56}+e^{78},\qquad & \eta_{1}\gamma_{(4)}\eta_{1} & =(j_{2}+\ii j_{1})\wedge(e^{5}+\ii e^{6})\wedge(e^{7}+\ii e^{8}),\\
-\ii\bar{\eta}_{2}\gamma_{(2)}\eta_{2} & =j_{1}-e^{58}+e^{67}, & \eta_{2}\gamma_{(4)}\eta_{2} & =(j_{3}+\ii j_{2})\wedge(-e^{5}+\ii e^{8})\wedge(e^{6}+\ii e^{7}),
\end{aligned}
\end{equation}
where the common $\SU 2$ structure is given by
\begin{equation}
j_{1}=e^{14}+e^{23},\qquad j_{2}=e^{13}+e^{42},\qquad j_{3}=e^{12}+e^{34},
\end{equation}
with $\theta_{\SU 2}=j_{2}+\ii j_{1}.$ This is the canonical $\SU 2$ structure that we use for calculations in the main text. 

Alternatively, the spinors can be thought of as defining a pair of canonical $\Gx 2$ structures which intersect on an $\SU 2$. Using the above unit-norm spinors, we again define the non-chiral Majorana spinors
\begin{equation}
\theta_{1}=\frac{1}{\sqrt{2}}(\epsilon_{1}+\epsilon_{2}),\qquad\theta_{2}=\frac{1}{\sqrt{2}}(\epsilon_{3}+\epsilon_{6}).
\end{equation}
In terms of these, the two $\Gx 2$ structures in eight dimensions are defined by
\begin{equation}
\begin{aligned}\theta_{1}\gamma_{(4)}\theta_{1} & =-\tfrac{1}{2}j_{3}^{2}-j_{3}\wedge e^{56}-j_{1}\wedge e^{67}+j_{2}\wedge e^{57},\\
\theta_{1}\gamma_{(3)}\gamma_{9}\theta_{1} & =j_{3}\wedge e^{7}+j_{1}\wedge e^{5}+j_{2}\wedge e^{6}+e^{567},\\
\theta_{1}\gamma_{(1)}\theta_{1} & =e^{8},\\
\theta_{2}\gamma_{(4)}\theta_{2} & =-\tfrac{1}{2}j_{3}^{2}-j_{3}\wedge e^{78}+j_{1}\wedge e^{58}-j_{2}\wedge e^{57},\\
\theta_{2}\gamma_{(3)}\gamma_{9}\theta_{2} & =j_{3}\wedge e^{5}+j_{1}\wedge e^{7}+j_{2}\wedge e^{8}+e^{578},\\
\theta_{2}\gamma_{(1)}\theta_{2} & =e^{6}.
\end{aligned}
\end{equation}
The common $\SU 2$ structure is then picked out by the pair of real one-forms
\begin{equation}
\theta_{2}\gamma_{(1)}\theta_{1}=\tfrac{1}{2}(e^{5}-e^{7}),\qquad\theta_{2}\gamma_{(1)}\gamma_{9}\theta_{1}=\tfrac{1}{2}(e^{5}+e^{7}).
\end{equation}

\subsection{A local \texorpdfstring{$\SU2$}{SU(2)} structure\label{subsec:A-local--1}}

We now want to relate the spinors that define the canonical $\SU 2$ structure above to the pair of non-chiral Majorana spinors $\chi_{i}$ that appear in the Killing spinor equations. Without loss of generality, we can write these in terms of the spinors which define the canonical $\SU 3$ and $\SU 2$ structures as
\begin{equation}
\begin{aligned}\chi_{1} & =\tfrac{1}{\sqrt{2}}(a_{1}\epsilon_{1}+a_{2}\epsilon_{2}),\\
\chi_{2} & =\tfrac{1}{\sqrt{4}}(b_{1}\epsilon_{1}+b_{2}\epsilon_{2}+b_{3}\epsilon_{3}+b_{4}\epsilon_{4}+b_{6}\epsilon_{6}),
\end{aligned}
\label{eq:KS_coeffs-1}
\end{equation}
where the coefficients $a_{i}$ and $b_{i}$ are real so that $\chi_{i}$ are Majorana. We see that setting $b_{6}=0$ recovers the local $\SU 3$ structure case of Appendix \ref{sec:An-orthonormal-frame}.

First, one can solve the normalisation and orthogonality constraints for the spinors:
\begin{equation}
\begin{aligned}\chi_{1}\chi_{1} & \equiv1=\tfrac{1}{2}(a_{1}^{2}+a_{2}^{2}),\\
\chi_{2}\chi_{2} & \equiv1=\tfrac{1}{5}(b_{1}^{2}+b_{2}^{2}+b_{3}^{2}+b_{4}^{2}+b_{6}^{2}),\\
\chi_{2}\chi_{1} & \equiv0=\tfrac{1}{\sqrt{10}}(a_{1}b_{1}+a_{2}b_{2}).
\end{aligned}
\label{eq:norm_ortho-1}
\end{equation}
  One finds
\begin{equation}
\begin{gathered}a_{2}=m\sqrt{2-a_{1}^{2}},\qquad b_{1}=-mn\frac{1}{\sqrt{2}}\sqrt{2-a_{1}^{2}}\sqrt{5-b_{3}^{2}-b_{4}^{2}-b_{6}^{2}}\\
b_{2}=n\frac{a_{1}}{\sqrt{2}}\sqrt{5-b_{3}^{2}-b_{4}^{2}-b_{6}^{2}},
\end{gathered}
\label{eq:orth_norm_conds-2}
\end{equation}
where $m^{2}=n^{2}=1$ encode the allowed choices of signs. Note that we need
\begin{equation}
a_{1}^{2}\leq2,\qquad b_{3}^{2}+b_{4}^{2}+b_{6}^{2}\leq5,\label{eq:ineqality-1}
\end{equation}
for a solution to exist. One can then match the remaining scalar bilinears with the parametrisation in terms of $\zeta$ and $S$ given in \eqref{eq:zeta_S_def}. Explicit expressions for the remaining bilinears can be found in the Mathematica notebook accompanying the submission.

\bibliographystyle{utphys}
\bibliography{draft,extra}

\end{document}